\documentclass[aps,prd,twocolumn,superscriptaddress,nofootinbib]{revtex4-2}

\usepackage{amsmath,amssymb}
\usepackage{graphicx}
\usepackage{booktabs}
\usepackage{multirow}
\usepackage{slashed}
\usepackage{xcolor}
\usepackage[colorlinks=true,linkcolor=blue!60!black,citecolor=blue!60!black,urlcolor=blue!60!black]{hyperref}

\newcommand{\alr}{A_{LR}}
\newcommand{\ecm}{E_{\rm cm}}
\newcommand{\kw}{(k_W)}
\newcommand{\kpp}{(k_{\phi\phi})}
\newcommand{\scf}{Sun-centered frame}

\begin{document}

\title{First search for sidereal and annual modulation of $Z$-pole asymmetries \\ with the recovered SLD data set}

\author{Wonyong Chung}
\affiliation{Department of Physics, Princeton University, Princeton, New Jersey 08544, USA}

\date{\today}

\begin{abstract}
We report what is, to our knowledge, the first time-resolved analysis of electroweak asymmetries at the $Z$ pole, using the public release of the recovered SLD collision data ($6.6\times10^{5}$ polarized $e^{+}e^{-}\to Z$ events from the 1996--1998 SLC runs; 385\,583 selected hadronic $Z$ decays spanning 782 sidereal cycles).
The left--right asymmetry $\alr$ is reconstructed with the release's implementation of the published SLD hadronic selection and the per-event signed Compton polarization and shows no sidereal, solar-diurnal, or annual modulation. A simultaneous unbinned likelihood fit of both daily templates with per-run-block nuisance parameters bounds the sidereal first-harmonic amplitude at $0.0086$ (95\%~C.L., statistical).
The analysis was developed and frozen on the initial partial release and applied unchanged to the 229\,833 events of the subsequently released 1998 run, confirming the
null hypothesis (amplitude $0.0018$, $p=0.90$) with the sensitivity following the expected
$\sqrt{P^{2}N}$ scaling.
Using the as-built SLC beam azimuth from the release's digitized documentation
corpus, the measurement bounds the anisotropy tensor of the asymmetry in the Sun-centered
frame, $\rho_{1}=\sqrt{\kappa_{XZ}^{2}+\kappa_{YZ}^{2}}\le0.060$,
$\rho_{2}=\sqrt{\tfrac14(\kappa_{XX}-\kappa_{YY})^{2}+\kappa_{XY}^{2}}\le0.18$,
and, for rank-one (vector) backgrounds, $\rho_{0}\le0.092$ (95\%~C.L.).
A leading-order matching computation maps the bounds onto minimal
Standard-Model-Extension coefficients as local linearized sensitivities at the
$10^{-2}$ level to the two-time-index block $\kw^{TJTK}$ of the electroweak
gauge-sector coefficient $k_{W}$, components on which no laboratory constraint
exists (previously reachable only through a one-sided astrophysical estimate of a trace
combination), and to the Higgs-sector coefficient $k_{\phi\phi}$. Because these
insertions split the $Z$ pole by polarization, for the tensor families assessed (one
eigenvalue pair $\pm\rho$, third eigenvalue zero) the resonance saturates the sidereal
amplitude at ${\approx}10^{-2}$, i.e. within the Born model the bounds exclude finite
coefficient bands rather than state an upper limit. The electron coefficient $d_{e}$ receives no profile limit at this statistics once the
closing-velocity factor and the sign of the ISR-dressed lineshape slope are included.
Additionally, we derive a previously unpublished observable from the helicity-odd azimuthal modulation of the thrust axis about the
beam -- a $Z$-pole \emph{birefringence channel} -- whose response to the same tensors
is enhanced by ${\approx}2M_Z/\Gamma_Z$ and is by construction invisible to every azimuth-integrated observable. A Born-level pilot of this channel on these events is null, and a dedicated analysis is in preparation.
We further report the first sidereal analysis of hadronization observables and of the
leptonic $Z$-pole asymmetries, bounding instrumental drifts at the
$10^{-4}$--$10^{-3}$ fractional level, providing a systematics benchmark for future
sidereal fragmentation analyses. To our knowledge these are the first time-resolved
and the first anisotropy constraints from $Z$-pole asymmetry data.
\end{abstract}

\maketitle

% ======================================================================
\section{Introduction}
\label{sec:intro}

Lorentz invariance underpins relativistic quantum field theory, and searches for its
violation are a rare experimental window on Planck-scale physics. The Standard-Model
Extension (SME)~\cite{Colladay:1996iz,Colladay:1998fq} organizes all such effects into an
effective field theory whose coefficient space has been mapped by two decades of
experiments~\cite{Kostelecky:2008ts}. The canonical laboratory signature is a
\emph{sidereal modulation}: a fixed background tensor in the \scf{}
\cite{Kostelecky:2002hh} is scanned by any Earth-fixed beam direction with the Earth's
rotation period $T_{\oplus}=23^{\rm h}56^{\rm m}04^{\rm s}$, imprinting harmonics of
$\omega_{\oplus}=2\pi/T_{\oplus}$ on any orientation-sensitive observable.

Collider tests of this signature fall into two classes. Sidereal modulation of
inclusive production rates has been searched for by D0 and CMS in top-quark pair
production~\cite{D0:2012rbu,CMS:2024tt} and by ZEUS in deep-inelastic
scattering~\cite{ZEUS:2022pbe}, while quark-sector SME frameworks have been developed for
deep-inelastic scattering and for gauge operators of arbitrary
dimension~\cite{Kostelecky:2016pyx,Lunghi:2018uwj,Kostelecky:2018yfa}, and $Z$-production
effects have been constrained at the LHC with a simulated sidereal analysis in
Ref.~\cite{Lunghi:2021jhep}.
Sidereal variation of neutral-meson oscillation parameters and their CPT-violating
flavor asymmetries has been searched for in charm (FOCUS~\cite{FOCUS:2003cpt}), beauty
(BaBar~\cite{BaBar:2008cpt}, D0~\cite{D0:2015cpt}, LHCb~\cite{LHCb:2016cpt}), and kaon
interferometry (KLOE~\cite{KLOE:2014cpt}). Nonetheless, the \emph{electroweak coupling asymmetries}
, i.e. the parity-violating precision observables of the $Z$ program, which measure
chiral couplings rather than rates or flavor mixing, have never been time-resolved.
In the electroweak gauge and Higgs sector of the SME, the current Data
Tables~\cite{Kostelecky:2008ts} (Table~D46) contain no entry derived from $Z$-pole
asymmetries, with the only entries from dedicated laboratory analyses being nuclear
$\beta$-decay orientation tests~\cite{Muller:2013wck,Sytema:2016xiv} (with
Refs.~\cite{NoordmansWilschutTimmermans:2013,Noordmans:2013oia} behind the derived
entries) and the existing $k_{W}$
entries are order-of-magnitude estimates from two combinations,
$\kw^{\alpha}{}_{X\alpha T}$ and $\kw^{\alpha}{}_{Y\alpha T}$, estimated from E158
within a three-parameter slice~\cite{Fu:2016fmf}, and a one-sided astrophysical
estimate of a trace combination~\cite{Altschul:2007kr}. (Decay-side SME effects on unpolarized
$Z\to\ell^{+}\ell^{-}$ observables were studied theoretically in
Ref.~\cite{Iltan:2005zll}.)
As Michel and Sher noted for the E158 estimate, ``the timing of events \ldots are not
available''~\cite{Michel:2019tti} -- the explicit time-dependence fit proposed in that work has never been performed on any polarized asymmetry.

This paper performs that fit at the $Z$ pole. The enabling development is the recovery
and public release of the SLD collision data~\cite{Cheng:2026sld,zenodo:v2}, which
provides per-event UTC timestamps, per-event signed Compton polarization, and
reproducible implementations of the published event selections. SLD remains the only
$Z$ factory with event-by-event initial-state helicity, and the SLC polarized source's
pulse-by-pulse (120~Hz) pseudorandom helicity reversal, designed to cancel slow
instrumental drifts in the DC asymmetry~\cite{SLD:2000jop}, is precisely the property
a time-resolved search needs, i.e. any detector or luminosity variation that is not
correlated with the helicity choice cancels identically in $\alr(t)$ at every
timescale. The left--right asymmetry is arguably the cleanest sidereal observable
available at any collider, and we find that the SLC collision-axis geometry turns out (Sec.~\ref{sec:geometry})
to be accidentally near-optimal for the classic first-harmonic SME components.

We distinguish two qualitatively distinct signal hypotheses by frequency: a fixed
cosmic frame produces modulation at the sidereal period, while couplings to the solar
gravitational environment produce solar-diurnal ($24^{\rm h}$) and annual periods. We fit
both daily templates simultaneously, and the annual phase separately (the 779-day span contains 2.1 sidereal--solar beat cycles, decohering the templates globally).

This analysis uses only the public release and is not an SLD Collaboration
result, and we note that the recovered event samples differ slightly from those of the original publications (Sec.~\ref{sec:dataset}). All bounds quoted are statistical at 95\%~C.L. and Sec.~\ref{sec:systematics} bounds the subleading systematic effects.

% ======================================================================
\section{Data set, event selection, and blind confirmation}
\label{sec:dataset}

The analysis uses the full 1996--1998 public release
(Zenodo~\href{https://doi.org/10.5281/zenodo.21199778}{10.5281/zenodo.21199778}), comprised of 68 mini-DST shards containing 658\,241 $Z$-pole events with per-event UTC timestamps
(\texttt{IEVENTH.evttime} validated against the run-block year labels) and per-event
signed Compton polarization (\texttt{PHBM.pol}; $\mathrm{pol}<0$ denotes a left-handed
electron beam, verified on data by reproducing the sign of the published asymmetry).
The data for the 1992--1995 runs have not as of yet been released publicly. The luminosity-weighted mean
center-of-mass energies were $\ecm=91.26\pm0.03$~GeV (1996) and $91.237\pm0.029$~GeV
(1997--98), calibrated by resonance scan~\cite{SLD:2000jop}.

The hadronic selection is the public release's reimplementation of the published 2000
$\alr$ selection~\cite{SLD:2000jop}, and we note that its energy-imbalance cut is evaluated as $|\sum p_z|/\sum|\mathbf{p}|$ over reconstructed particles rather than as the magnitude of the calorimeter energy-vector sum over the scalar energy sum of the original publication; the other cuts follow the published definitions. In total, 385\,583 events with valid polarization ($|P_e|>0.1$) and timestamps spanning 1996-04-20 to 1998-06-08 (779.4 days $=$ 781.6 sidereal cycles), with $\langle|P_e|\rangle=0.735$
(0.730 in 1998, matching the run history). Leptonic selections follow the published
2001 criteria~\cite{SLD:2000ldu}: 16\,547 $ee$, 13\,992 $\mu\mu$, and 17\,133
$\tau\tau$ candidates.

Table~\ref{tab:dc} compares the raw DC
asymmetry with the published measured values at the running energy, epoch by
epoch, as opposed to the pole-extrapolated $A_{LR}^{0}$. The recovered 1996
sample reproduces the published value (difference $0.0005$, $0.06$ of the quadrature
error) and the recovered 1997--98 sample reproduces the published measured 1997--98
asymmetry $0.1454\pm0.0024$ (difference $0.0009$, $0.3$ of the quadrature error). The
samples overlap heavily but are not identical, and the raw values predate the
$[+0.16\pm0.07]\%$ relative correction of Eq.~(2) of Ref.~\cite{SLD:2000jop}
(backgrounds, positron polarization, energy asymmetry), so we attach no significance to either difference.
For a modulation search, we note that DC offsets are absorbed exactly by $A_0$ and the block nuisances of Sec.~\ref{sec:method}.

\begin{table}[tb]
\caption{DC benchmarks of the recovered samples against the published measured
values at the running energy (statistical errors; the pole-extrapolated values,
$0.15929$ and $0.14906$, are not the comparators). The recovered values are raw
counting asymmetries divided by the mean polarization, while published values include the
small corrections of Ref.~\cite{SLD:2000jop}. NB: the samples overlap but are not
identical.}
\label{tab:dc}
\begin{ruledtabular}
\begin{tabular}{lcc}
Sample & This analysis (raw) & Published~\cite{SLD:2000jop} \\
\colrule
1996 ($\ecm=91.26$)     & $0.1554\pm0.0058$   & $0.1559\pm0.0057$ \\
1997--98                & $0.1445\pm0.0024$   & $0.1454\pm0.0024$ \\
1998 only (blind stage) & $0.14553\pm0.00284$ & --- \\
1996--1998 combined     & $0.14602\pm0.00218$ & --- \\
\end{tabular}
\end{ruledtabular}
\end{table}

The analysis was constructed as a blind measurement by pure coincidence of the data set's release schedule: the selection, phase conventions, binning, fit machinery, and look-elsewhere accounting (i.e. the entire chain) was developed and frozen on the initial (v1) release which contained only the 1996 and partial 1997 runs (155\,748 selected hadronic $Z$). When the v2 data set was released, newly released events were identified against the v1 release as being the complete 1998 run, 229\,833 selected events, $1.48\times$ the entire development sample. Two v1 shards were also re-released in v2 with corrected timestamp events that were corrupted in v1.

% ======================================================================
\section{Method}
\label{sec:method}

For each event we compute the local mean sidereal and solar phases at SLAC
(longitude $-122.2025^{\circ}$) from the UTC timestamp, using the standard GMST
polynomial~\cite{Meeus:1998}; a phase-convention offset rotates the harmonic
coefficients $(C,S)$ into each other but leaves every amplitude and $p$-value
invariant. Since the raw asymmetry is $\mathcal{O}(P\alr)\simeq0.107$, the per-event
information is nearly Gaussian and two complementary estimators are used.

\emph{Binned estimator.} In each of 8 phase bins,
$\alr = (N_L-N_R)/(N_L+N_R)/\langle|P_e|\rangle_{\rm bin}$, with nested weighted
least-squares fits (constant; $+$first; $+$second harmonic) and likelihood-ratio
$p$-values.

\emph{Joint unbinned likelihood.} The reported statistic is a simultaneous fit of both
daily templates plus per-run-block nuisance offsets. For event $i$ with helicity
$h_i=\pm1$ ($+$ for left), polarization magnitude $|P_i|$, and run block $b(i)$,
\begin{align}
\alr(t_i) &= A_0 + \beta_{b(i)}
 + C_s\cos\phi_{\rm sid}(t_i) + S_s\sin\phi_{\rm sid}(t_i) \nonumber\\
 &\quad + C_d\cos\phi_{\rm sol}(t_i) + S_d\sin\phi_{\rm sol}(t_i),
\label{eq:model}
\end{align}
with the exact binomial likelihood
$P(h_i) = \tfrac12\!\left[1 + h_i |P_i| \alr(t_i)\right]$
(the 120~Hz pseudorandom reversal makes the helicity sequence an unbiased coin flip),
16 block offsets $\beta_b$ constrained to zero weighted mean, Fisher-matrix errors, and
Wilks likelihood-ratio tests. The blocks, defined by the mini-DST run-group boundaries, absorb slow drifts and genuine epoch-to-epoch asymmetry differences; because each block spans weeks to months of running, the block profile is nearly orthogonal to the daily harmonics. The per-event $|P_i|$ from the Compton polarimeter enters the likelihood directly, so polarization
variations at any timescale are absorbed event-by-event. Over the full data set the
sidereal and solar templates decohere (2.1 beat cycles). The residual template
correlations in the joint fit are $\le0.30$ and are propagated by the Fisher matrix.

\emph{Calibration.} The $p$-value calibration of both estimators was verified in two ways with scrambled-phase permutation ensembles (1000 binned, 100 joint; the
modulation-$p$ distributions are consistent with uniform, Kolmogorov--Smirnov
$p=0.80$ and $0.13$), and, because the permutation null severs the
block-to-helicity association that the fitted null retains, with a parameterization of the fitted composite null (observed times, blocks, and polarizations
kept exactly, helicities redrawn from the fitted per-event probabilities). The
parameterization confirms the Wilks calibration for each reported statistic separately
(200 replicates per unbinned statistic, 1000 for the binned; $6.0\%$ and $5.8\%$ of
replicates below $p=0.05$ for the headline and binned statistics), with the final values
KS $p=0.17$ (headline no-sidereal test), $0.96$ (no-solar), $0.62$ (four-parameter
omnibus), and $0.48$ (binned estimator).

\emph{Signal response.} The binned estimator's signal response is not unity:
averaging a harmonic of order $m$ over the $2\pi/8$-wide bins multiplies its fitted
amplitude by $\mathrm{sinc}(m\pi/8)$, with $0.9745$ for $m{=}1$, $0.9003$ for $m{=}2$.
This response is measured directly by a signal-injection method run through the full
binning-and-fit chain (synthetic, deterministic real-exposure, and stochastic
end-to-end injections) and the phase-averaged real-exposure gains agree with the analytic
factor to better than $10^{-3}$, with a phase-to-phase scatter below $0.3\%$.
Binned amplitudes are quoted as raw estimator outputs - every binned
quantity used as a limit is divided by the response, which also divides
the associated uncertainty so that the limit construction retains its coverage. The
joint likelihood is unbinned and its response is unity by the same
injection test for both harmonic orders, and modulation $p$-values are
response-independent throughout.

Hadronization observables (mean thrust $\langle T\rangle$, mean charged multiplicity
$\langle n_{\rm ch}\rangle$, mean thrust-axis polar angle $\langle|\cos\theta_T|\rangle$) are analyzed in two channels: the raw means, which are \emph{not} protected by the
helicity reversal and therefore track instrumental drifts; and the
helicity-correlated differences $\langle O\rangle_L-\langle O\rangle_R$, which inherit
the full reversal protection. We note the thrust axis throughout is the release's implementation, which maximizes $\sum_i|\mathbf{p}_i\cdot\hat n|$ over candidate axes along single momenta and pair sums, i.e. a restricted-axis approximation to the exact thrust maximization, expected to be close for two-jet-dominated hadronic $Z$ events but not evaluated against an exact algorithm here.

% ======================================================================
\section{Beam orientation and geometry}
\label{sec:geometry}

Interpreting the harmonics requires the beam direction. For a horizontal beam at
geodetic latitude $\lambda$ and azimuth $\alpha$ (E of true N, along the $e^-$ motion),
the beam unit vector in the \scf{} has constant declination
$\sin\delta = \cos\lambda\,\cos\alpha$ and precesses about the celestial pole once per
sidereal day.

The SLC collision axis is fixed by the as-built survey drawing in the release's
documentation corpus (CN-316~\cite{Friedsam:1985cn316}), calibrated to its printed
benchmark coordinates and measured against the SLAC site grid, whose east axis is
the linac; the linac bearing itself is survey-grade, $261.45^{\circ}$ up-beam from a
1963 Coast \& Geodetic Survey determination quoted in SLAC alignment
documentation~\cite{LCLS:2003tn} and confirmed to $0.01^{\circ}$ by the georeferenced
3-km building footprint. The final-focus control line through the interaction region
then gives an $e^{-}$ heading of
\begin{equation}
\alpha = 166.0^{\circ}\pm0.8^{\circ},
\label{eq:azimuth}
\end{equation}
$5.4^\circ$ from perpendicular to the linac, corroborated independently by the
printed final-focus dipole budget (CN-296~\cite{Sheppard:1985cn}), by the SLD
convention notes ($+z$ = $e^{+}$ direction $\approx$ North), and by the published
End Station A site bearing of Ref.~\cite{Fu:2016fmf}. (The arcs' \emph{total} bend,
$236^{\circ}$/$226^{\circ}$ per arc including reverse-bend regions, is not usable as
geometry input, as the net turns are $+84.4^{\circ}$/$-95.6^{\circ}$.) A geodetic tie
beyond $\pm0.8^{\circ}$ is not publicly reachable as the documents fixing the
final-focus geometry exactly were not distributed, and no NGS control mark carries
SLC-grid coordinates. With $\lambda=37.42^{\circ}$,
\begin{gather}
\delta = -50.4^{\circ},\qquad |\sin2\delta| = 0.982\pm0.002,\nonumber\\
\cos^{2}\delta = 0.406\pm0.004 .
\end{gather}
The near-north--south collision axis at mid-latitude places $|\sin2\delta|$ within 2\%
of its maximum: by accident of its site-constrained arcs, SLD is a near-optimal geometry
for first-harmonic ($\kappa_{XZ},\kappa_{YZ}$) sidereal searches, precisely the classic
SME target components (Fig.~\ref{fig:kappa}, right). The residual geometry uncertainties
contribute $\pm0.2\%$ ($\rho_1$) and $\pm1\%$ ($\rho_2$), negligible against
statistics.

% ======================================================================
\section{Results}
\label{sec:results}

\begin{figure*}[tb]
\includegraphics[width=0.98\textwidth]{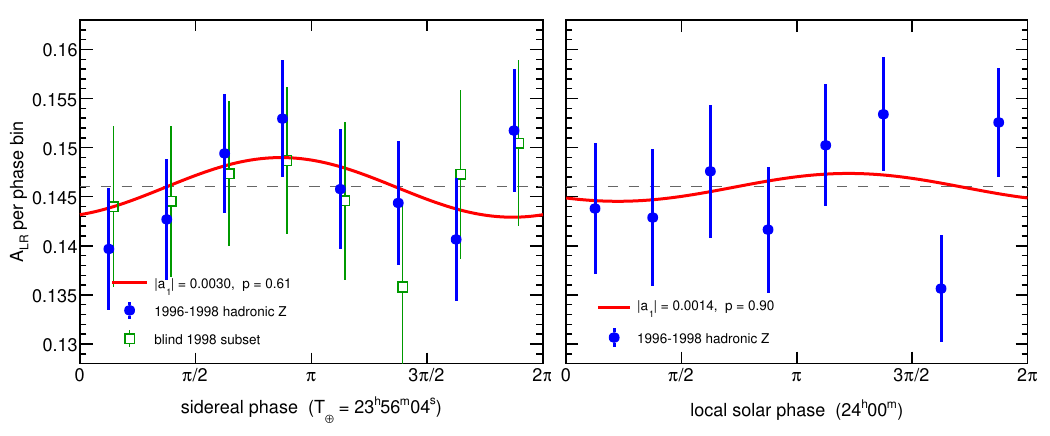}
\caption{$\alr$ versus sidereal (left) and local solar (right) phase for the combined
1996--1998 hadronic sample (filled circles), with the first-harmonic fit (solid) and the
constant fit (dashed). Open squares (left panel): the 1998 blind subset, offset for
visibility. All modulation amplitudes are consistent with zero.}
\label{fig:phase}
\end{figure*}

\emph{Blind confirmation (1998).} The frozen analysis applied to the 229\,833 unseen
events report DC $\alr = 0.14553\pm0.00284$, consistent with both the development
sample and the published 1997--98 value, and the sidereal first harmonic $|a_1| = 0.0018$ with
$p=0.90$; solar $|a_1|=0.0018$, $p=0.91$. Second harmonics and the scrambled control report null. The per-coefficient sensitivity $\sigma(C_1)=0.0040$ matched the
$\sqrt{\langle P^{2}\rangle N}$ scaling expectation from the development sample
($0.0048\times\sqrt{86{\rm k}/122{\rm k}}=0.0040$, with $86$k and $122$k the
effective statistics $N\langle P\rangle^{2}$ of the development and unseen samples.

\begin{figure}[tb]
\includegraphics[width=0.48\textwidth]{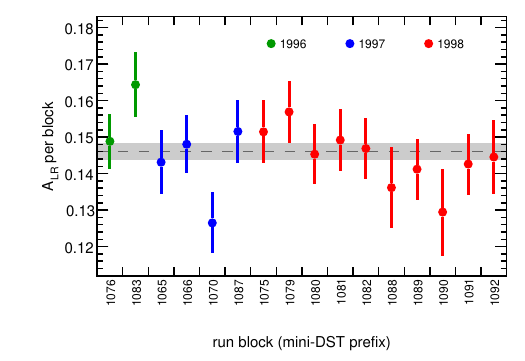}
\caption{$\alr$ per run block (16 mini-DST groups, 1996--1998). The band is the
combined statistical error. The block offsets are nuisance parameters in the joint fit.}
\label{fig:blocks}
\end{figure}

\begin{figure*}[tb]
\includegraphics[width=0.98\textwidth]{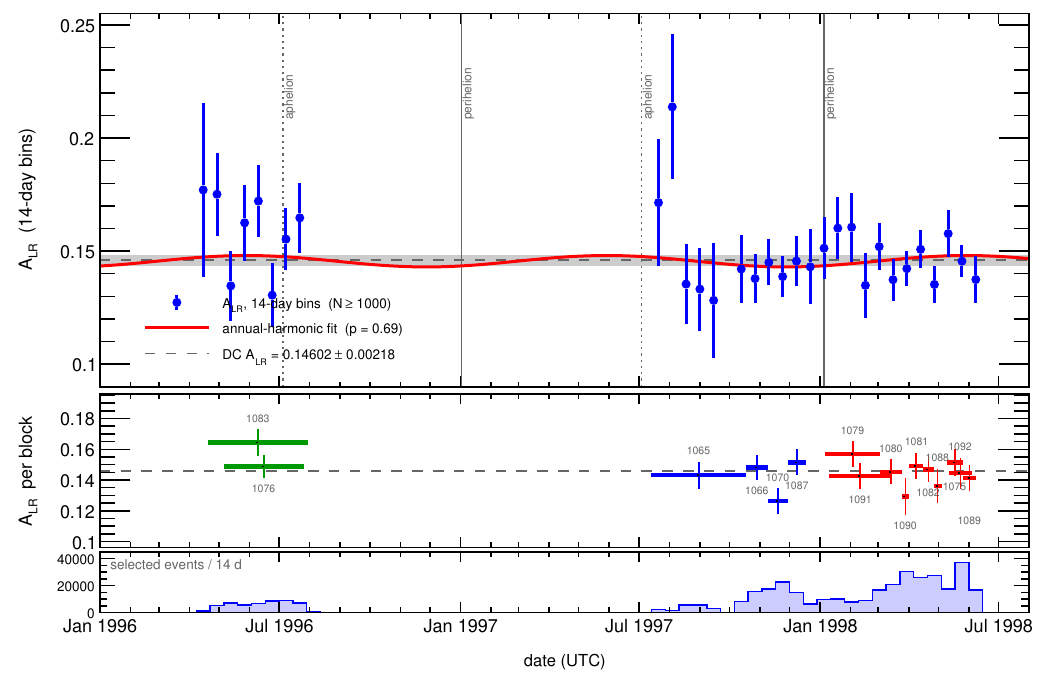}
\caption{$\alr$ versus calendar time (UTC), 1996--1998. Top: $\alr$ in 14-day
bins (bins with $N \ge 1000$ selected events), with the DC value
$\alr = 0.14602 \pm 0.00218$ (dashed line; the band is the combined statistical
error) and the annual-harmonic fit of Sec.~\ref{sec:results} (solid curve,
$p=0.69$); vertical lines mark the perihelia (solid) and aphelia (dotted).
Middle: $\alr$ per run block, each drawn over the block's true time span,
colored by year as in Fig.~\ref{fig:blocks} (labels: mini-DST group numbers).
The blocks are interleaved file groups, not disjoint time epochs (several
1998 spans overlap) and the block offsets are nuisance parameters in the
joint fit [Eq.~(\ref{eq:model})]. Bottom: selected events per 14-day bin,
matching the binning of the top panel and showing the exposure gap between
August 1996 and July 1997.}
\label{fig:time}
\end{figure*}

\begin{figure}[tb]
\includegraphics[width=0.48\textwidth]{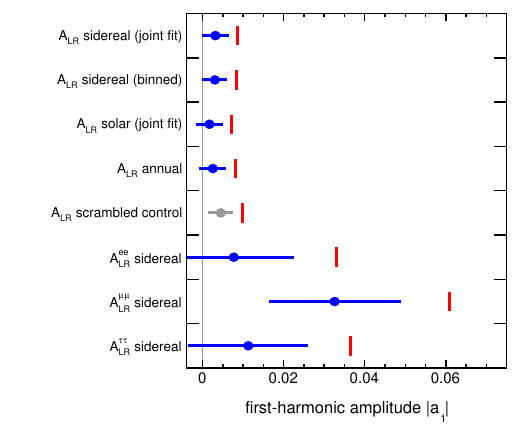}
\caption{First-harmonic amplitudes across channels. Points: fitted amplitudes with
the larger of the two coefficient uncertainties as the error bar; red ticks:
$|a_1|+1.645\max(\sigma_C,\sigma_S)$, the diagonal form of the 95\%~C.L. construction of
Sec.~\ref{sec:results} (the two forms differ by less than $0.3\%$ for the sidereal rows
and the scrambled control, by about $1\%$ for the solar rows and by $5.6\%$ for the
annual row), divided by the
binned-estimator response $0.9745$ for the binned rows (Sec.~\ref{sec:method}); the
joint-fit rows have unit response.}
\label{fig:forest}
\end{figure}

\begin{figure*}[tb]
\includegraphics[width=0.85\textwidth]{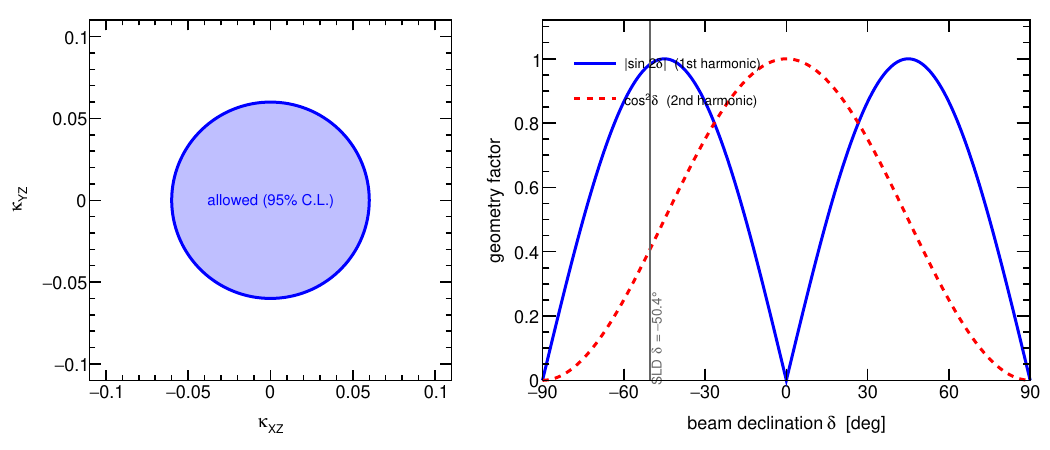}
\caption{Left: allowed region for the first-harmonic tensor components. Right: the
geometry factors versus beam declination; the SLC collision axis sits within 2\% of the
first-harmonic optimum.}
\label{fig:kappa}
\end{figure*}

\emph{Combined harmonics.} Table~\ref{tab:harmonics} gives the combined first-harmonic
results, Fig.~\ref{fig:phase} shows the phase distributions, Fig.~\ref{fig:blocks}
the block stability, and Fig.~\ref{fig:time} the time structure of the sample. Every
channel is null. The joint-fit 95\%~C.L. upper limit on the
sidereal first-harmonic amplitude is
\begin{equation}
a_{1}^{\rm sid} \;\le\; 0.0086 \qquad (95\%~\text{C.L., stat}),
\label{eq:ul}
\end{equation}
constructed as $|\hat a_1| + 1.645\sqrt{\lambda_{\max}(V)}$, with
$\lambda_{\max}$ the larger eigenvalue of the fitted $(C,S)$ covariance $V$, a
construction whose coverage is $\ge95\%$ for every true amplitude and orientation,
exact in the Gaussian limit (verified numerically on the measured covariances).
For the fits that set quoted limits the $C$--$S$ correlations are negligible
($-1\times10^{-4}$ for the joint sidereal first harmonic, $-0.002$ for the binned sidereal
second harmonic, and $0.010$ for the binned sidereal first harmonic), so $\sqrt{\lambda_{\max}}$
coincides with $\max(\sigma_{C},\sigma_{S})$ to better than $0.2\%$ and every quoted limit
is unchanged at the reported precision. The solar and annual fits, whose non-uniform
exposure correlates the two coefficients at $0.04$ (solar, binned and joint) and $0.20$
(annual), set no quoted limit. Their ticks in Fig.~\ref{fig:forest} use the diagonal form
$\max(\sigma_{C},\sigma_{S})$, which for the annual channel reads $0.0082$ against
$0.0087$ with $\sqrt{\lambda_{\max}}$.

The sidereal second harmonic is a binned-only quantity in the frozen analysis (the
frozen joint model carries first harmonics only). Dividing the binned result
[$\hat a_2 = 0.0046$, $\sigma\simeq0.0031$ per coefficient] by the $m{=}2$
response $0.9003$ (Sec.~\ref{sec:method}) gives
\begin{equation}
a_{2}^{\rm sid} \;\le\; 0.0107 \qquad (95\%~\text{C.L., stat}),
\label{eq:ul2}
\end{equation}
with the same method applied to the response-corrected estimate (without the
correction the analysis would undercover for true amplitudes above
${\sim}3\sigma$). As a decorrelating cross-check, a post-unblinding extension
of the joint likelihood with second-harmonic templates at both daily frequencies (unbinned, unit response by the injection test) finds all four added coefficients
consistent with zero (sidereal and solar second-harmonic $p=0.72$ and $0.30$; both
together, $p=0.61$; the first-harmonic coefficients shift by $\le0.13\sigma$) and
bounds $a_{2}^{\rm sid}\le0.0080$, tighter than Eq.~(\ref{eq:ul2}). We quote the
corrected binned bound as the conservative choice. The largest block pulls
($-0.019\pm0.008$ and $+0.018\pm0.009$) are the 1997 and 1996 outliers already
present in the development sample. The ten 1998 blocks are unremarkable
($\chi^{2}=5.3/9$ within 1998).

\begin{table}[tb]
\caption{First-harmonic fits to $\alr(t)$, combined 1996--1998 (385\,583 events).
``Joint'' rows are from the simultaneous unbinned likelihood
[Eq.~(\ref{eq:model})]; the anti-sidereal control is a single-template unbinned fit
with the same block nuisances (Sec.~\ref{sec:systematics}); $p$ is the modulation
$p$-value (Wilks for the unbinned fits; for the binned fits the fit's
$\Delta\chi^{2}$ between the constant and constant-plus-harmonic models, which for
the non-uniform annual exposure differs from a diagonal Wald statistic built from
the printed coefficients). Binned amplitudes are raw estimator
outputs with signal response $0.9745$ (Sec.~\ref{sec:method}); joint amplitudes
have unit response; $p$-values are response-independent.}
\label{tab:harmonics}
\begin{ruledtabular}
\begin{tabular}{lcccc}
Template & $C_1$ & $S_1$ & $|a_1|$ & $p$ \\
\colrule
Sidereal (binned) & $-0.0028(31)$ & $+0.0012(31)$ & $0.0030$ & $0.61$ \\
Sidereal (joint)  & $-0.0031(32)$ & $+0.0007(33)$ & $0.0032$ & $0.62$ \\
Solar (binned)    & $-0.0011(31)$ & $-0.0009(31)$ & $0.0014$ & $0.90$ \\
Solar (joint)     & $-0.0007(32)$ & $-0.0015(33)$ & $0.0017$ & $0.88$ \\
Annual            & $-0.0021(32)$ & $+0.0014(33)$ & $0.0025$ & $0.69$ \\
Scrambled control & $+0.0044(31)$ & $-0.0010(31)$ & $0.0045$ & $0.34$ \\
Anti-sidereal control & $-0.0003(31)$ & $+0.0014(31)$ & $0.0014$ & $0.90$ \\
\end{tabular}
\end{ruledtabular}
\end{table}

\emph{Annual channel.} With all eight perihelion-phase bins populated across the
779-day span (populations spanning a factor 9.6), the annual first harmonic is
$0.0025$ ($p=0.69$) -- to our knowledge the first annual-modulation search on any
electroweak precision observable. An untargeted broadband periodicity scan of the CPT
parameter $z$ in $B^{0}$ mixing at BaBar spanned $0.26\,\mathrm{yr}^{-1}$ to
$2.1\,\mathrm{day}^{-1}$, a band that contains the annual frequency, without a
dedicated annual analysis or limit~\cite{BaBar:2008cpt}. Outside colliders, CPLEAR
bounded differential gravitational couplings of the neutral-kaon system against the
annual variation of the solar potential and the monthly variation of the lunar
potential~\cite{Apostolakis:1999cplear},
and LEP energy-calibration records have been reinterpreted at two Earth--Sun distances
to bound the gravitational mass of the positron~\cite{Kalaydzhyan:2016positron}.
We note that for this analysis, run blocks span
weeks to months and are therefore partially degenerate with an annual template within
a two-year window, so, unlike the daily fits, the quoted fit uses no block nuisances
and slow drifts are not deducted. An event-level robustness study re-fits the channel under three drift models of increasing conservatism (an unconstrained linear drift, free per-block offsets, and free offsets for each of the 16 populated calendar months) and the result is null under all three ($p = 0.67$,
$0.82$, $0.78$), with the per-coefficient uncertainty inflating from $+9\%$ to
$3.6\times$ ($\sigma\simeq0.012$) to ${\simeq}7\times$ ($\sigma\simeq0.022$--$0.026$)
as the drift model is freed. The channel is therefore a search with quantified drift sensitivity, not a precision bound.

\emph{Equivalence-principle reading.} In general relativity the ambient solar
potential has no intrinsic local effect: the detector, the beams, and the laboratory
fall freely together, and the residual tidal terms, of relative size
$GM_\odot L^{2}/(r^{3}c^{2})\approx4\times10^{-29}$ across a 10\,m apparatus, are
negligible. A perihelion-locked coefficient $C_{1}$ in excess of the conventional
seasonal effects bounded above would therefore signal a dependence of local
electroweak physics on the gravitational potential, i.e., a violation of local
position invariance. Writing
$\delta\sin^{2}\theta_{W}^{\rm eff}/\sin^{2}\theta_{W}^{\rm eff}=k\,\delta\Phi/c^{2}$,
the annual harmonic amplitude of the solar potential is
$e\,GM_\odot/(a c^{2})=1.6\times10^{-10}$ ($3.3\times10^{-10}$ peak to peak), and
with $|dA_{LR}/d\sin^{2}\theta_{W}^{\rm eff}|\approx7.9$ the annual covariance
envelope ($0.0087$; Fig.~\ref{fig:forest} displays the diagonal form, $0.0082$)
corresponds to a sensitivity
$|k|\lesssim2.9\times10^{7}$ (statistical; ${\approx}1.9\times10^{8}$ under the most
conservative drift model above) in the sense of this
section, not a quoted limit. Clock comparisons bound the analogous
electromagnetic-sector coefficient at the $10^{-6}$--$10^{-7}$
level~\cite{Safronova:2018rev}, reaching $1.4(1.1)\times10^{-8}$ from the
$^{171}$Yb$^{+}$ E3/E2 frequency ratio~\cite{Lange:2021lpi}, while CPLEAR bounded differential kaon--antikaon
couplings to the annual variation of the solar potential at the $10^{-9}$
level~\cite{Apostolakis:1999cplear}, and E\"otv\"os-type experiments indirectly
constrain spatial variation of the weak mixing angle through its contribution to
nuclear binding~\cite{Kraiselburd:2012eo}. The present channel is, to our knowledge,
the first direct annual search with a $Z$-pole coupling asymmetry. Countershaded
matter--gravity couplings~\cite{Kostelecky:2011matter}, which evade
non-gravitational laboratory searches, motivate $\Phi$-tracking signals of this
kind, though no matching from those coefficients to
$\sin^{2}\theta_{W}^{\rm eff}$ is attempted here (at leading order in the orbital
eccentricity the template is a cosine of perihelion phase with the amplitude as its
only free parameter).

\emph{Auxiliary channels.} Table~\ref{tab:aux} collects the hadronization and
leptonic results. The unprotected event-shape means bound instrumental drifts at the
$10^{-4}$--$10^{-3}$ fractional level at \emph{both} daily frequencies, with no
significant modulation anywhere, providing a systematics benchmark for future sidereal
fragmentation analyses at facilities without helicity reversal. The flip-protected
differences and the three leptonic asymmetries are null across the board. The
development-sample fluctuations all regressed on the full statistics as expected from the look-elsewhere control: the $\mu\mu$ excursion ($p=0.025$ on 1996--97
data) to $p=0.135$ at $2.5\times$ statistics, the thrust-mean excursions (sidereal
$p=0.009$, solar $0.015$) to $0.44$ and $0.055$, and the deepest excursion in the development sample, the $\langle|\cos\theta_T|\rangle$ protected difference at solar phase
($p=0.0011$), regressed to $0.11$ (Sec.~\ref{sec:systematics}).

\begin{table}[tb]
\caption{Auxiliary channels, combined 1996--1998. Event-shape means are fractional
modulation amplitudes at both daily frequencies (unprotected by the helicity
reversal, i.e. they track instrumental drifts and serve as the drift benchmark, and
modulation $p$-values across the mean channels span 0.055--0.95, with the thrust
mean shown as the worked example in the text). Protected differences
$\langle O\rangle_L-\langle O\rangle_R$ and the leptonic asymmetries inherit the
full reversal protection, and their sidereal fits are listed ($p$-values; the
protected thrust difference is bounded at $8.7\times10^{-4}$, 95\%~C.L.,
response-corrected). The corresponding solar-channel fits belong to the same
21-test suite (Sec.~\ref{sec:systematics}) and are all null.}
\label{tab:aux}
\begin{ruledtabular}
\footnotesize\setlength{\tabcolsep}{3pt}
\begin{tabular}{lccc}
Channel & $A_0$ & Sidereal & Solar \\
\colrule
\multicolumn{4}{l}{\emph{Mean fractional amplitudes (unprotected)}} \\
$\langle T\rangle$ & --- & $2.1\times10^{-4}$ & $4.0\times10^{-4}$ \\
$\langle n_{\rm ch}\rangle$ & --- & $1.1\times10^{-3}$ & $8.9\times10^{-4}$ \\
$\langle|\cos\theta_T|\rangle$ & --- & $4.0\times10^{-4}$ & $2.4\times10^{-3}$ \\
\colrule
\multicolumn{4}{l}{\emph{Flip-protected differences ($p$)}} \\
$\langle T\rangle_{L-R}$ & --- & 0.50 & --- \\
$\langle n_{\rm ch}\rangle_{L-R}$ & --- & 0.067 & --- \\
$\langle|\cos\theta_T|\rangle_{L-R}$ & --- & 0.098 & --- \\
\colrule
\multicolumn{4}{l}{\emph{Leptonic $\alr$ (amplitude;~$p$)}} \\
$ee$ & $0.1496\pm0.0105$ & 0.008;~0.87 & --- \\
$\mu\mu$ & $0.1546\pm0.0114$ & 0.033;~0.135 & --- \\
$\tau\tau$ & $0.1267\pm0.0104$ & 0.011;~0.74 & --- \\
\end{tabular}
\end{ruledtabular}
\end{table}

% ======================================================================
\section{Anisotropy bounds}
\label{sec:bounds}

Any leading-order rotation-violating, CPT-even modification of a scalar observable
measured along a beam direction $\hat b$ can be written
\begin{equation}
O(t) = \bar O\left[\,1 + \kappa_O^{JK}\,\hat b_J(t)\,\hat b_K(t)\,\right],
\label{eq:kappa}
\end{equation}
with $\kappa_O$ a constant symmetric traceless tensor in the \scf{}. Inserting the
geometry of Sec.~\ref{sec:geometry}, the harmonic amplitudes measure the
rotation-invariant combinations
\begin{align}
a_1 &= \bar A\,|\sin2\delta|\;\rho_1, &
\rho_1 &\equiv \sqrt{\kappa_{XZ}^{2}+\kappa_{YZ}^{2}}, \nonumber\\
a_2 &= \bar A\,\cos^{2}\delta\;\rho_2, &
\rho_2 &\equiv \sqrt{\tfrac14(\kappa_{XX}-\kappa_{YY})^{2}+\kappa_{XY}^{2}},
\end{align}
while a rank-one (vector) background $\kappa^{J}\hat b_J$ produces a first harmonic
with \emph{no} second harmonic, i.e. a rank discriminator the measurement provides for
free, with $a_1 = \bar A\cos\delta\,\rho_0$, $\rho_0\equiv\sqrt{\kappa_X^2+\kappa_Y^2}$.
With $\bar A = 0.14602$ and the geometry factors above, Eqs.~(\ref{eq:ul}) and
(\ref{eq:ul2}) give \emph{the first bounds on the anisotropy tensor of any
$Z$-pole asymmetry}:
\begin{equation}
\boxed{\;\rho_1 \le 0.060, \qquad \rho_2 \le 0.180, \qquad \rho_0 \le 0.092\;}
\label{eq:rhobounds}
\end{equation}
(95\%~C.L., statistical; $\rho_2$ includes the binned-estimator response
correction of Sec.~\ref{sec:method}, and the unbinned cross-check would give
$\rho_2\le0.135$). These are deliberately quoted at the observable level: any
rank-two SME coefficient maps onto $\kappa^{JK}$ with an operator-specific response
that is linear only at small coefficient values (Sec.~\ref{sec:matching}. For the
propagator channels the linearization holds to 5\% only for $\rho\lesssim0.01$), so
future theory work can reuse the bounds without re-touching the data provided it
carries the operator's full response, not a linear factor. The amplitude bounds are
independent of the local-sidereal-time phase convention
and of the azimuth sign, only $|\delta|$ enters.

% ======================================================================
\section{SME operator matching}
\label{sec:matching}

We now map the observable-level bounds onto minimal-SME coefficients. The complete
leading-order computation (exact tree-level trace algebra with modified spinors,
vertices, dispersion relations, and the resonant propagator expanded to first order in
each coefficient) is summarized in Appendix~\ref{app:matching}, here we give the
structure and the results. Three paths feed $\alr(t)$ at the pole, separated
by the ratio structure of the asymmetry:

\emph{(i) Chirality filter.} A common rescaling of both chiral couplings,
$x_{L,R}\to x_{L,R}(1+\epsilon)$, cancels in $\alr$ identically. The
chirality-symmetric electron coefficient $c_{e}$ therefore decouples at
$\mathcal{O}(1)$ and survives only through the beam-dispersion channel (iii); the
chirality-odd coefficient $d_{e}$ (with $c_{L,R}=c\pm d$) enters unsuppressed through
the modified vertex and spinors, amplified by the chiral enhancement
$\partial\alr/\partial\ln(g_L/g_R) = 1-\bar A^{2}$ relative to the coupling shift, i.e.\
by $(1-\bar A^{2})/\bar A\simeq6.7$ relative to $\bar A$ itself.

\emph{(ii) Additive shifts: the $\gamma$--$Z$ mixing channel.} After electroweak
symmetry breaking, the gauge-sector coefficient $k_{W}$ generates a $\gamma$--$Z$
kinetic mixing insertion $-2\tan\theta_W \kw_{\kappa\lambda\mu\nu}q^{\kappa}q^{\mu}$
\cite{Fu:2016fmf,Michel:2019tti} (with $k_B=-\tan^2\theta_W\,k_W$ imposed by the
photon-sector constraints). This adds a helicity-\emph{independent}
term to the effective $Z$ coupling on the electron side. The asymmetry is invariant under common rescalings but not common shifts: a shift $\delta g$ enters as
$\delta\alr = 2[(x_L-x_R) - \bar A\,(x_L+x_R)]\,\delta g/(x_L^2+x_R^2)$ with the
full $1/\bar A$ enhancement. Because the $Z$ four-momentum is purely timelike in the
collision frame, the contraction $\kw^{\kappa\lambda\mu\nu}\hat q_\kappa \hat q_\mu$
projects exactly the two-time-index block $\kw^{T\lambda T\nu}$, orthogonal, component
by component, to the one-time-index slice $\kw^{\alpha}{}_{J\alpha T}$ estimated from
E158~\cite{Fu:2016fmf}. The only prior information on these components is a starred,
one-sided astrophysical estimate of the trace combination
$\kw^{\alpha}{}_{J\alpha K} = \kw_{TJTK} - \sum_l \kw_{lJlK}$, in which they appear
summed against the fully spatial block~\cite{Altschul:2007kr}. No laboratory bound of
any kind exists. The same $k_W$ also enters through the $ZZ$-kinetic insertion
$-2(1-\tan^{2}\theta_W)\kw qq$, an effective $Z$-mass tensor
$\kpp^{\rm eff} = -2(1-\tan^{2}\theta_W)(s/M_Z^{2})\kw^{T\lambda T\nu}$ that partially
cancels the mixing channel (Appendix~\ref{app:matching}). The transverse electron
currents contract the
spatial block $\kw^{TjTk}$ into precisely the rank-two form of Eq.~(\ref{eq:kappa}).

\emph{(iii) Lineshape-slope channels.} Coefficients that shift the effective
center-of-mass energy or the $Z$ mass tensor direction-dependently (i.e. $c_{e}$ through
the beam dispersion at fixed machine settings, $\kpp$ and the $ZZ$-kinetic part of
$k_W$ through the resonance denominator) enter through the $s$-dependence of the
asymmetry ($d\alr/d\sqrt{s} = +0.022$~GeV$^{-1}$ Born, consistent with the
ZFITTER-based correction used by SLD~\cite{SLD:2000jop}) and, for chirality-odd
dispersion shifts, through the full lineshape slope, for which we use SLD's quoted
energy-asymmetry coefficient, $4.3\pm2.9$ at $\ecm=91.237$~GeV ($2.0\pm3.0$ at
$91.26$~GeV)~\cite{SLD:2000jop}, with the $\pm2.9$ dominated by the $\pm29$~MeV
energy-scale knowledge. This value is the
logarithmic slope $\ecm\,\sigma'/\sigma$ of the observed ISR-dressed cross
section, and it is positive: radiative return shifts the observed peak to
${\approx}91.28$~GeV, so the dressed cross section rises at $91.237$~GeV, whereas
every Born lineshape falls there ($\ecm\sigma'/\sigma\simeq-7.8$). Integrating the $\mathcal{O}(\alpha)$-exponentiated radiator over the
$s$-dependent-width Breit-Wigner
gives $+4.11$ at $91.237$~GeV and $+1.81$ at $91.26$~GeV, matching the SLD results in magnitude, trend, and uncertainty. With the $\gamma$--$Z$ interference and a
running width the numbers are $+4.24$ and $+1.94$. The ISR dressing of the $d_e$
dispersion channel is therefore derived from the methods used for SLD corrections.
The propagator channels ($\kpp$ and the $ZZ$-kinetic part of $k_W$) are
not dressed by this input as their matching factors below are Born-level.
Because the measured asymmetry is a ratio of radiatively convolved rates, their
ISR-dressed derivative is $\partial_K\alr^{\rm ISR} = \langle\partial_K\alr\rangle +
\mathrm{Cov}(\alr,\partial_K\ln\Sigma)$ with $\Sigma=\sigma_L+\sigma_R$ and
$\langle f\rangle \equiv \int\! H\,\Sigma\,f \,/\! \int\! H\,\Sigma$ being the average over the radiator-weighted rate with the covariance term in the same. The
total ISR change is $-8.5\%$ of the Born factor for $\kpp$, dominated by the covariance term, and $-8.6\%$ for the
$ZZ$-kinetic part of $k_W$, netting $+5.3\%$ on the $k_W$ total, whose mixing part changes only at the
$10^{-3}$ level. The
selected-event value is not computed -- the factors are quoted at Born level with this
stated dressing.

The explicit computation (Appendix~\ref{app:matching}) yields the modulating
structures, linear in each coefficient,
\begin{align}
\delta\alr(t) &= -1.116\,\kw^{T\hat b T\hat b}(t)
\;\;[\,{-1.815}\,\text{mix}+0.699\,ZZ\,] \nonumber\\
 &= -0.499\,\kpp^{\hat b\hat b}(t) \;=\; +\beta\, d_e^{\hat b\hat b}(t),
\nonumber
\end{align}
each stable to $<\!0.1\%$ across the $\pm29$~MeV energy-scale knowledge (Born level
for the propagator channels, as stated above). For $d_e$ the dispersion-free part is
$\beta_0 = 2\beta_{\rm vtx} = 1.955$, with $\beta_{\rm vtx} = 0.9775$ the archived value
of the trace computation at $91.238$~GeV: the vertex and modified spinors
contribute $1-A^{2}$ exactly, and the per-crossing yield contributes another $1-A^{2}$
through the helicity-dependent closing velocity of the two beams, i.e. the overlap
integral of two counter-propagating bunches cancels the $(v_1+v_2)$ of the luminosity
definition, not the $1/v_{\rm rel}$ of the cross section, and the same chiral
coefficient acts on both members of the annihilating pair, so the left- and
right-handed channels close at $2(1\mp d)$ (Appendix~\ref{app:matching}). In the
fixed-momentum machine convention (the SLC case, since the arcs and energy
spectrometers define the beams through magnetic rigidity) the beam-dispersion
channel adds
$\beta_{\rm vtx}[-\ecm\sigma'/\sigma] + \bar A\,\ecm\,(d\alr/d\ecm) = -3.91\pm2.83$
with SLD's positive dressed slope, giving $\beta = -1.96\pm2.83$, and the two terms nearly
cancel, and $\beta$ is consistent with zero ($0.7\sigma$). A profile likelihood over
$(\rho_d,\varphi,\beta)$ with $\beta$ as a Gaussian nuisance then returns no limit -- decoupling ($\beta\to0$) costs $\Delta\chi^{2}=0.48$, below both the one-sided 90\%
($1.64$) and 95\% ($2.71$) thresholds, so we conclude \emph{the $d_e$ profile in the
fixed-momentum convention yields no limit at this statistics}, and no $d_e$ row is
quoted. The fixed-energy scenario ($\beta=\beta_0$, no dispersion channel) would give
envelopes of $4.5\times10^{-3}$ (first harmonic) and $1.35\times10^{-2}$ (second). These are conditional scenario values, not limits.

Following the Data-Tables convention~\cite{Kostelecky:2008ts}, each row of
Table~\ref{tab:sme} treats the quoted coefficient with all other SME coefficients set
to zero. The linear factors above are \emph{local} sensitivities, and their range of
validity is finite. The $\kpp$ and $ZZ$-kinetic insertions enter the $Z$ propagator
as polarization-dependent pole shifts, $G^{-1} = (s-M_Z^{2}+iM_Z\Gamma_Z)\,\mathbb{1}
+ M_Z^{2}K$ in the eigenbasis of the spatial tensor $K$ (for $k_W$, $K$ is the
$ZZ$-kinetic effective tensor above, and the $\gamma$--$Z$ mixing enters the same
$2\times2$ neutral propagator), so the expansion parameter is $M_Z\rho/\Gamma_Z =
36.5\rho$ for $\kpp$ and $51.2\rho$ for $k_W$, which is of order one at
$\rho\sim10^{-2}$. We therefore evaluate the harmonic amplitudes with the propagator
inverted exactly, for the one-parameter tensor family of each row. For these families the resonance that gives a
$Z$-pole search its sensitivity also caps the sidereal amplitude: the
exact amplitude rises linearly, turns over at $M_Z\rho/\Gamma_Z\sim1$, and falls,
with maxima of $0.0100$--$0.0161$ (first harmonic) and $0.0062$--$0.0098$ (second)
against the measured envelopes $0.0086$ and $0.0107$. Consequently the
second-harmonic envelope excludes \emph{no} value of $\rho$ in either sector, and the
first-harmonic envelope excludes finite bands, $0.023\le\rho\le0.073$ for $\kpp$ and
$0.0085\le\rho\le0.105$ for $k_W$, rather than everything above a limit. At the
linear-sensitivity values themselves the exact-to-linear ratios are $0.84$/$0.57$
($\kpp$, $m=1$/$2$) and $0.93$/$0.76$ ($k_W$), and the linearization holds to 5\%
only for $\rho\lesssim0.009$ ($\kpp$) and $0.007$ ($k_W$) in the first harmonic
($0.014$ and $0.010$ in the second). Table~\ref{tab:sme} therefore reports, for each
row, the linear sensitivity, the saturation maximum, and the set actually excluded
within this model; the model-independent statement remains Eq.~(\ref{eq:rhobounds}).
These are statements about the specified tensor families in the paper's Born,
inclusive, ISR-free model, not a maximum over arbitrary simultaneous tensors, i.e. a
family with no zero eigenvalue (eigenvalues $(\rho,\rho,-2\rho)$, say) does not turn
over in this model, because every polarization's pole leaves $\sqrt s$ and the
photon exchange takes over, and such large displacements of the resonance are outside
the scope of a linear analysis of the asymmetry -- and they use no information from
other observables on the same coefficients (the $Z$ lineshape itself constrains a
polarization-split pole; Sec.~\ref{sec:conclusion}).
The $\kw^{TJTK}$ rows are, to our knowledge, the first laboratory
\emph{sensitivity} to these four combinations --- the Data Tables carry no laboratory
entry for them --- but they are not bounds; the $\kpp$ rows are sensitivities
obtained at the $Z$ resonance, far from competitive with the indirect low-energy
bounds on related coefficient combinations~\cite{Kostelecky:2008ts,Anderson:2004qi},
reported for completeness of the coefficient-space map because they probe the weak
vertex directly with an independent systematic basis.

CPT-odd \emph{fermion} coefficients do not contribute:
$b_{e}$ effects are helicity-suppressed by $m_e/E_b\simeq1.1\times10^{-5}$, and
constant vector backgrounds decouple from chirality-diagonal processes. The CPT-odd
gauge terms $k_{1,2}$ are set to zero in this analysis: their photon-sector
combination $k_{AF}$ is bounded far below any collider reach~\cite{Kostelecky:2008ts},
and although the remaining $Z$-sector combination can propagate with positive
energies below a bound and be quantized consistently~\cite{Colladay:2017zen} (the
photon-sector positivity argument of Ref.~\cite{Colladay:1998fq} does not exclude it),
it would populate the rank-one channel through an anisotropic $\gamma$--$Z$
Chern--Simons mixing at a level already constrained by cosmic-ray
estimates~\cite{Kostelecky:2008ts}, and we do not treat that channel here. The
antisymmetric $k_{\phi W}, k_{\phi B}$ enter after symmetry breaking only as constant
background field strengths, with no tree-level $s$-channel funnel; boost-mixed tensor
components cancel between the two symmetric beams. With $k_{1,2}=0$ the rank-one
bound $\rho_0$ has no leading occupant in the minimal SME, and a rank-one detection
would have pointed to that CPT-odd $Z$-sector channel or beyond the minimal framework
(Appendix~\ref{app:matching}). All statements here are for the minimal, $d\le4$ SME;
nonminimal operators generate the same and higher harmonics with additional energy
weighting and are outside our scope.

\begin{table*}[tb]
\caption{Minimal-SME coefficient sensitivities in the Sun-centered frame, one
coefficient family at a time (all other SME coefficients zero); no row is an upper
limit. ``Linear'' is the local linearized sensitivity: the 95\%~C.L. amplitude
envelope [Eqs.~(\ref{eq:ul}), (\ref{eq:ul2}); the second harmonic response-corrected,
Sec.~\ref{sec:method}] divided by the linear matching and geometry factors.
``Maximum'' is the largest harmonic amplitude the family can produce when the
resonant propagator is inverted exactly, at the coefficient value in parentheses.
``Excluded'' is the set of $\rho$ whose exact amplitude exceeds the measured
envelope ($0.0086$ for $m=1$, $0.0107$ for $m=2$); against the unbinned second-harmonic
cross-check $a_2\le0.0080$ the $k_W$ second harmonic would exclude $0.022$--$0.087$.
The $d_e$ profile yields no limit (text). ``Funnel'': the $\gamma$--$Z$ mixing plus
$ZZ$-kinetic insertions for $k_W$, the $\gamma$--$Z$ interference for $\kpp$. ``Prior''
is the status in the Data Tables~\cite{Kostelecky:2008ts}, Table D46, read in full
($^{*}$: starred/derived estimates, no dedicated laboratory analysis --- for $k_W$ a
one-sided astrophysical estimate of a trace combination; for $\kpp$ indirect bounds
through the electron $c$ coefficients~\cite{Anderson:2004qi} and starred entries
derived from nuclear $\beta$-decay
data~\cite{NoordmansWilschutTimmermans:2013,Noordmans:2013oia} at
$5\times10^{-7}$--$10^{-6}$ on the very components of these rows).}
\label{tab:sme}
\begin{ruledtabular}
\begin{tabular}{lcccccc}
Combination $\rho$ & Funnel & $m$ & Linear & Maximum (at $\rho$) & Excluded & Prior \\
\colrule
$\sqrt{(\kw^{TXTZ})^2+(\kw^{TYTZ})^2}$ & mixing $+$ $ZZ$ & 1 & $7.8\times10^{-3}$ & $0.0161$ ($0.029$) & $0.0085$--$0.105$ & trace est.$^{*}$ \\
$\sqrt{\tfrac14(\kw^{TXTX}\!-\!\kw^{TYTY})^2+(\kw^{TXTY})^2}$ & mixing $+$ $ZZ$ & 2 & $2.4\times10^{-2}$ & $0.0098$ ($0.044$) & none & trace est.$^{*}$ \\
$\sqrt{(\kpp^{XZ})^2+(\kpp^{YZ})^2}$ & interference & 1 & $1.8\times10^{-2}$ & $0.0100$ ($0.041$) & $0.023$--$0.073$ & indirect$^{*}$ \\
$\sqrt{\tfrac14(\kpp^{XX}\!-\!\kpp^{YY})^2+(\kpp^{XY})^2}$ & interference & 2 & $5.3\times10^{-2}$ & $0.0062$ ($0.061$) & none & indirect$^{*}$ \\
\end{tabular}
\end{ruledtabular}
\end{table*}

\subsection{A helicity-odd azimuthal channel of the final-state axis}
\label{sec:birefringence}

The same events carry a second, independent channel to the same tensors. For a
spatial tensor insertion $K$ in the $Z$ propagator (the $\kpp$ insertion above, or
the $ZZ$-kinetic part of $k_W$), the linear response of $\alr$ resolved in the
\emph{unsigned} final-state axis $\hat n$ about the beam $\hat b$ is, by explicit
evaluation with helicity spinors in the Born model,
\begin{equation}
\delta\alr(\hat b,\hat n) = \alpha\,\frac{(\hat n\times K\hat n)\cdot\hat b}{1+c^{2}}
 + \beta\,\frac{{\rm Tr}[P_{\hat b}KP_{\hat n}]}{1+c^{2}},\qquad c=\hat b\cdot\hat n,
\label{eq:biref}
\end{equation}
with $P_{\hat u} = \mathbb{1}-\hat u\hat u^{\rm T}$ and $|\alpha| =
2(M_Z/\Gamma_Z)(1-A_e^{2}) = 71.4$ at the pole ($Z$ exchange only); evaluated at
$\sqrt s = 91.238$~GeV with the $\gamma$--$Z$ interference the model gives
$|\alpha| = 71.2$ per unit $\kpp$ component and $99.8$ per unit $\kw^{TJTK}$ component
through the $ZZ$-kinetic insertion (the mixing insertion contributes $0.05\%$ of it),
and $\beta = 0.9964$, whose average over $\hat n$ is
$\tfrac12\beta({\rm Tr}K - K^{\hat b\hat b}) = 0.4982\,({\rm Tr}K - K^{\hat b\hat b})$,
the beam-only factor of the matching computation above ($0.4986$ there, in the
single-counted photon convention of the archived code; $0.1\%$ apart). The mechanism is two lines long: the $Z$
produced by an electron of helicity $h$ is circularly polarized, $\hat x + ih\hat y$
about the beam; a spatial $K$ places the poles of the two transverse linear
polarizations at $M_Z^{2}(1\mp\kappa)$, so on resonance their propagators acquire a
relative phase $2M_Z\kappa/\Gamma_Z$, converting the circular polarization into a
helicity-dependent elliptical one whose orientation the decay axis measures as a
$\sin2\phi$ pattern. It is the $Z$-sector analogue of the photon-sector vacuum
birefringence that is the canonical SME observable~\cite{Kostelecky:2002hh}, with the
resonance supplying the lever $M_Z/\Gamma_Z$. The $\alpha$ term is odd in the beam
helicity and vanishes under azimuthal averaging: it is invisible to every
$\phi$-integrated observable, including the beam-only analysis of this paper, and the
helicity reversal protects it against helicity-\emph{independent} acceptance effects
in the same way as the rate.

A Born-level pilot on the 385\,583 thrust axes of this analysis (the release's thrust
implementation, Sec.~\ref{sec:method}), in the documented SLD
detector frame ($+z$ along the positron beam, $+y$ vertical, $+x$ west, right-handed;
$\phi_T$ right-handed about $+z$, so that the azimuth about the electron direction
$\hat b=-\hat z$ is $-\phi_T$), with this
paper's block nuisances and the four daily beam templates of
Eq.~(\ref{eq:model}) as free parameters and the five Sun-centered-frame components of
$K$ as signal templates through Eq.~(\ref{eq:biref}), is null: all five components are
consistent with zero (Wilks $p=0.63$ for the five-parameter test; the largest is the
static component, $|\hat\kappa_{ZZ}| = 2.3\pm1.6\times10^{-4}$ for $\kpp$, $1.5\sigma$;
the beam templates return their values of Table~\ref{tab:harmonics}). The Born
statistical 95\% envelopes are $(2.8,\,2.0,\,4.9)\times10^{-4}$ on
$(\rho_1^{n},\rho_2^{n},|\kappa_{ZZ}|)$
for $\kpp$ and $(2.0,\,1.4,\,3.5)\times10^{-4}$ for $k_W$, before any dilution --- here
$\kappa$ are the Sun-centered-frame components of $\kpp^{JK}$ itself and, for $k_W$,
of $\kw^{TJTK}$ itself, which enters Eq.~(\ref{eq:biref}) through the $ZZ$-kinetic
effective tensor $-2(1-\tan^{2}\theta_W)(s/M_Z^{2})\kw^{TJTK}$ (hence the coefficient
$99.8$ per unit $\kw^{TJTK}$ component, and envelopes $1.40$ times tighter
than the $\kpp$ ones); $\rho_1^{n}$ and $\rho_2^{n}$ are the combinations defined in
Sec.~\ref{sec:bounds}, with the superscript marking the final-state-axis channel,
and $\kappa_{ZZ}$ is the celestial-pole component the beam-only analysis cannot see.
Unlike in Table~\ref{tab:sme}, the linearization is valid at these values
($M_Z\rho/\Gamma_Z\approx0.01$--$0.02$; exact-to-linear $0.99$); the envelopes use
the construction of Sec.~\ref{sec:results} on the fitted $2\times2$ covariance
blocks; and the ISR dressing of $\alpha$, computed as the derivative of the convolved
azimuth-resolved rates, is $-4.5\%$. We report the laboratory-frame static
coefficient alongside the $p$-value rather than behind it: the coefficient of
$\sin^{2}\theta_T\sin2\phi_T/(1+\cos^{2}\theta_T)$ in $\alr$ is $a_s =
-0.0090\pm0.0050$ ($1.8\sigma$; the even $\cos2\phi_T$ partner is $-0.0024\pm0.0050$),
because a static helicity-odd laboratory pattern is exactly the signature of a
helicity-correlated angular acceptance and must be resolved before any static
component is interpreted. The sign of $\kappa$ relative to the SME operator convention
is not fixed by the pilot; we quote $|\alpha|$.

The pilot is Born-matched and statistical and is not a bound. A dedicated study with injection through a selected-event
response fitted with a helicity-dependent angular-efficiency nuisance model is in preparation. The
four sidereally modulated Sun-frame components are protected against a static
instrumental fake by the same anti-sidereal sideband logic as the rate channel
(Sec.~\ref{sec:systematics}), whereas the static $m=0$ component ($\kappa_{ZZ}$) is not, and
its bound is conditional on SLD's left--right beam-asymmetry records or on a
spin-state split that the public release does not carry. The closest studies in the literature are polarization-split $Z$ dispersion relations that have been solved for the CPT-odd
coefficients~\cite{Colladay:2017zen}, and split poles derived for a CPT-even kinetic
insertion~\cite{Kepuladze:2026lzg}, both without an azimuthal observable. KLOE's neutral-kaon test of CPT and Lorentz symmetry used azimuthal subdivisions of
the kaon direction about the beam together with sidereal-time bins~\cite{KLOE:2014cpt};
an azimuthal correlation that vanishes at Born level in the Standard Model and is
switched on by Lorentz violation was studied for polarized top
decay~\cite{Ebrahimi:2025hjx,Rezaei:2025wiw}; and SLD measured helicity-resolved
asymmetries of the three-jet event-plane orientation angle $\chi$ --- the angle
between the event plane and the quark--electron plane, with a separately defined
inclusive variant~\cite{SLD:2000ern,Burrows:1997dh} --- a different angle, in a QCD
and $T$-odd context. To our knowledge no $Z$-pole measurement has resolved a final-state axis
harmonically about the beam as a resonance-enhanced Lorentz test. The indirect bounds
on the electron $c$ coefficients that constrain $\kpp$ through field
redefinitions~\cite{Anderson:2004qi}, and the starred $\beta$-decay-derived entries on
the $\kpp$ spatial components themselves at $5\times10^{-7}$--$10^{-6}$
\cite{Kostelecky:2008ts,NoordmansWilschutTimmermans:2013,Noordmans:2013oia}, are many
orders of magnitude tighter than
any direct collider sensitivity, so the channel's value for $\kpp$ is as a direct
laboratory sensitivity; for the $\kw^{TJTK}$ block no laboratory constraint exists.

% ======================================================================
\section{Systematic considerations}
\label{sec:systematics}

The bounds of Eq.~(\ref{eq:rhobounds}) are statistical; this section argues that every
known systematic source of \emph{apparent} daily modulation is subdominant to
$\sigma(a_1)\simeq0.003$. Table~\ref{tab:syst} collects the magnitude budget.

\emph{Helicity-reversal protection.} Any \emph{multiplicative common-mode} effect
uncorrelated with the pseudorandom 120~Hz helicity choice (detector efficiency,
luminosity scale, deadtime, temperature, reconstruction drifts) cancels identically
in $\alr(t)$. This is the same property that protected the DC
measurement~\cite{SLD:2000jop}, applied per phase bin. The drift bounds from the
unprotected means (Table~\ref{tab:aux}) show its scale: effects of
$10^{-4}$--$10^{-3}$ that would contaminate an unprotected observable are absent from
the protected channels. Effects that fall outside this protection are additive or
helicity-correlated, and each is bounded in Table~\ref{tab:syst} by SLD's own
measurement of its DC magnitude~\cite{SLD:2000jop}, taken term by term from Eq.~(2)
of that reference rather than through the terms' signed sum.

\begin{table}[tb]
\caption{Magnitude budget for effects outside the helicity-reversal protection. Each
row bounds the apparent first-harmonic amplitude $a_1$ that the effect could produce
\emph{if} its modulated part does not exceed its DC-bounded magnitude
(see text for the status of that assumption); where an uncertainty is quoted, the
allowance is taken at one standard deviation. Entries in parentheses are instead the
instrumental excursion that would be required to fake $a_1=0.003$;
$\sigma(a_1)\simeq0.003$. DC knowledge from Ref.~\cite{SLD:2000jop}; the individual
conversions are given in the text.}
\label{tab:syst}
\begin{ruledtabular}
\footnotesize\setlength{\tabcolsep}{3pt}
\begin{tabular}{lll}
Effect & DC knowledge & Max fake $a_1$ \\
\colrule
Background dilution & $f_{\rm bkg}\sim10^{-3}$ & $2\times10^{-4}$ \\
L--R luminosity & $10^{-4}$ (monitors) & $1.4\times10^{-4}$ \\
L--R energy & $(+2.8\pm1.4)\times10^{-7}$ & $<10^{-5}$ \\
L--R polarization & $(+2.8\pm6.9)\times10^{-3}$ & $1\times10^{-4}$ \\
$e^+e^-$ background & $(0.013\pm0.013)\%$ & $2\times10^{-5}$ \\
Positron polarization & $(-0.02\pm0.07)\%$ & $7\times10^{-4}$ \\
Polarimeter scale & $0.5\%$ total & ($2\%$ needed) \\
$\ecm$ drift (common) & $+0.022$~GeV$^{-1}$ slope & (140~MeV needed) \\
\end{tabular}
\end{ruledtabular}
\end{table}

We comment on two rows of Table~\ref{tab:syst}: the
luminosity row, an additive raw-asymmetry offset $A_{\mathcal L}(t)$ is
helicity-correlated by construction and enters $\alr$ as
$A_{\mathcal L}/\langle P\rangle$; SLD's beam monitors bounded it at the $10^{-4}$
level DC (cross-checked at the $10^{-3}$ level by the luminosity-monitor Bhabha
asymmetry). A DC budget does not by itself bound a zero-mean modulation: an offset
$\ell(t)=\epsilon[\cos\phi_{\rm sid}(t)-\langle\cos\phi_{\rm sid}\rangle]$ has zero
exposure-weighted mean and an arbitrary sidereal amplitude, and since the raw
helicity asymmetry is exactly $q=(P\alr+\ell)/(1+P\alr\ell)$, an $\ell$-amplitude of
$2.4\times10^{-3}$ would produce an apparent $a_1=0.0032$ at $\langle P\rangle=0.735$
with the DC asymmetry unchanged. That the modulated part does not exceed the
DC-bounded magnitude is therefore an assumption about the machine, not a
measurement; machine intensity cycles are solar-clocked, with their seasonal
sidebands covered by the sideband analysis below. Second, the polarization rows: the
per-event Compton $|P_i|$ enters the likelihood directly, so helicity-independent
polarization variations are absorbed at any timescale. A helicity-\emph{correlated}
magnitude difference $A_P$ is not: the release's per-event polarization value is shared
by both helicity states (99.6\% of events carry a value that also occurs under the other
setting, 93\% within the same ten-minute window, and successive same-value events change
setting at the pseudorandom rate), so the polarization-asymmetry row treats $A_P$ as a
hidden magnitude difference with SLD's correction as its transfer, and equal effective
exposure conditional on polarization and time is assumed, not verified. A within-block
consistency test of that exposure model on the release --- the contrast
$C = N^{-1}\sum_i h_i\,(|P_i| - B_{b(i)})$ with $B_b = \sum_b|P|^{2}/\sum_b|P|$, which has
zero mean when the helicity choice is independent of the recorded polarization within a
block --- returns $-1.05\pm0.34\times10^{-4}$ ($3.1\sigma$), a static block-level feature
whose cause is not identified (a 1\% polarimeter resolution would account for about 7\%); as a polarization-correlated exposure imbalance it would allow at most
$1.4\times10^{-4}$ on the sidereal first harmonic, the size of the luminosity row. Therefore the values in Table~\ref{tab:syst} should be understood as allowances, not as corrections to the
reported amplitudes. The remaining systematic for the polarimeter is a hypothetical
sidereal-correlated \emph{scale} error, and faking $a_1=0.003$ would
require a coherent $2\%$ diurnal-sidereal gain oscillation against a total scale
uncertainty of $\sim0.5\%$ dominated by time-independent effects, with polarimeter
data taken continuously. Residual transverse polarization couples only through
azimuthal asymmetries, which vanish in the azimuth-integrated counting. A sidereal
modulation of $A_E$ has no known mechanism (accelerator thermal and operational
cycles follow solar time), and faking $a_1=0.003$ by common-mode energy drift would
require a sidereal-coherent 140~MeV oscillation of $\ecm$, over an order of magnitude
beyond plausible machine excursions; slow drifts are absorbed by the block nuisances.
The individual conversions behind Table~\ref{tab:syst} are that the energy asymmetry couples
through the steep lineshape, $\delta\alr = -\ecm\,[\sigma'(\ecm)/\sigma(\ecm)]\,
A_E/\langle P_e\rangle$ [Eq.~(2) of Ref.~\cite{SLD:2000jop}], a DC correction of
$2\times10^{-6}$; the polarization asymmetry enters as
$A_m^{2}A_P/\langle P_e\rangle$ under the hidden-difference model above (a per-event
likelihood whose recorded $|P_i|$ itself differed between the two helicity states would
instead respond with $(1-A_m^{2})/\langle P_e\rangle\approx1.35$ per unit $A_P$; the
shared per-event value excludes that model in its simple form); the background dilution
uses SLD's published fraction for its own selection, assumed to carry over to the
reimplemented one (the DC agreement of Table~\ref{tab:dc} supports the sample, not the
fraction); the $e^+e^-$ background asymmetry is
$0.023\pm0.022$; the efficiency asymmetry is negligible by the symmetric acceptance;
and the positron polarization enters $\alr$ additively, so its $7\times10^{-4}$
allowance is one standard deviation of the direct 1998 measurement.

\emph{Sidereal sidebands (aliasing).} The sidereal frequency is the upper annual
sideband of the solar-diurnal frequency
($\omega_{\rm sid}=\omega_{\rm sol}+\Omega_{\rm ann}$), so a solar-daily systematic
whose amplitude is seasonally modulated feeds the sidereal template with half
its modulation depth even when its constant solar-daily part is null; over this span the upper sideband is numerically
indistinguishable from the sidereal template itself, so no template fit can separate
them. The magnitude bounds above cover this channel under the same assumption (a
sideband cannot exceed its parent effect), and it is tested with a standard anti-sidereal control template
($\omega_{\rm sol}-\Omega_{\rm ann}$, period $24^{\rm h}03^{\rm m}57^{\rm s}$), which
no inertial-frame physics can populate, is null (joint amplitude
$0.0014$, $p=0.90$, at the full per-coefficient sensitivity; a control channel,
excluded from the signal-test suite by construction, like the scrambled control). An
annually modulated diurnal effect feeds the anti-sidereal band with the same amplitude
as the sidereal band when its two solar-diurnal quadratures share a common annual
phase; in general the sidereal and anti-sidereal sideband powers differ by
$C_sS_c - C_cS_s$ for quadrature envelopes $(C_0,C_c,C_s)$ and $(S_0,S_c,S_s)$, so a
single-sideband systematic is not excluded by this control alone.
The physics interpretation of the sidereal null is therefore conditional on the
absence of a star-locked, helicity-correlated instrumental modulation of amplitude
$\gtrsim2\times10^{-3}$ in the raw asymmetry. The public release
cannot test this assumption since its per-event banks carry no scaler, luminosity-monitor, or
Bhabha-rate time series, and we note the measured amplitudes themselves do not depend on it.
The second-harmonic bounds inherit the same logic one order up: the $m{=}2$
sidereal line is the \emph{second} annual sideband of the diurnal second harmonic,
and its mirror, the second harmonic of the anti-sidereal phase, can be fed by such
leakage (with equal power when the two quadratures share a common semiannual
phase) while remaining closed to inertial-frame physics. It is null (binned
amplitude $0.0040$, $p=0.42$; unbinned $0.0052$, $p=0.24$) and is a necessary control, not
a sufficient one.

\emph{Timing.} Timestamps are Unix UTC (validated against run-block year labels);
the two corrupted timestamps found in v1 were repaired upstream in v2, confirming the
integrity of the timestamp record in both directions. A uniform clock offset or GMST convention error rotates $(C,S)$
without changing amplitudes; sub-second jitter is irrelevant against the 86164-s period.
Phase coverage is excellent, and the largest-to-smallest phase-bin
population ratio is 1.14 sidereal and 9.6 annual, and the block nuisances absorb the
coverage-weighting of slow drifts.

\emph{Look-elsewhere.} The full suite comprises 21 first-harmonic tests ($\alr$
sidereal/solar/annual; three event-shape means and three protected differences at two
frequencies; three lepton channels at two frequencies). A programmatic enumeration of
the frozen suite gives the development-sample minimum $p=0.0011$ (the
$\langle|\cos\theta_T|\rangle$ protected difference at solar phase): under the global
null, treating the tests as independent, the probability of a value this small
arising somewhere in the suite is $1-(1-0.0011)^{21}\approx2.3\%$. It regressed to
$p=0.11$ in the combined data, as did every other sub-0.05 development excursion
($2.5\times$ statistics; Sec.~\ref{sec:results}). The combined sample's own suite
minimum is $p=0.055$, globally unremarkable ($\approx$70\% under the same accounting;
shared events correlate the tests, so both global values are approximations).

\emph{Geometry.} The azimuth band $\pm0.8^{\circ}$ contributes $\pm0.2\%$ to $\rho_1$
and $\pm1\%$ to $\rho_2$. The final-focus vertical slope and crossing angles are
$\mathcal{O}(\text{mrad})$ and negligible.

% ======================================================================
\section{Conclusion}
\label{sec:conclusion}

Using the recovered SLD data set, we performed the first time-resolved analysis of a
$Z$-pole asymmetry: a blind-confirmed null search for sidereal, solar,
and annual modulation of $\alr$ over 782 sidereal cycles, yielding the first
anisotropy-tensor bounds for any $Z$-pole observable [Eq.~(\ref{eq:rhobounds})] and,
through leading-order matching, the first laboratory sensitivity to four combinations
of the two-time-index gauge-sector block $\kw^{TJTK}$ and a $Z$-resonance sensitivity
to $\kpp$ --- sensitivities, not bounds, because for the tensor families studied the
polarization-split resonance saturates the sidereal amplitude and the envelopes
exclude finite bands (Table~\ref{tab:sme}); the $d_e$ profile yields no limit at this
statistics. The measurement demonstrates a
methodology --- helicity-reversal-protected time-resolved asymmetries, two-frequency
discrimination, internal initial-state controls, blind confirmation on later-released
data --- that transfers directly to the EIC~\cite{Accardi:2012qut}
and to any future polarized collider, where the same fit costs nothing beyond
timestamps. Within the recovered data, the remaining headroom is modest (the 1992--95
runs would add $\sim20\%$ in statistics if recovered); the qualitatively new target is a
sidereal analysis of the LEP lineshape archives, where mass-type coefficients enter
with resonance enhancement rather than through the asymmetry ratio --- a direction
whose theory side is developing~\cite{Jejelava:2026liv}. The same insertions split
the $Z$ pole for the three polarizations at every collider, so the unpolarized
lineshape is a composite of three Breit--Wigners and a profiled fit to the LEP
hadronic cross sections is the natural complement to the saturated sidereal channel
(the toy studies archived with this analysis establish sensitivity, not an
exclusion). Within this data set, the helicity-odd azimuthal channel of
Sec.~\ref{sec:birefringence}, with its ${\approx}2M_Z/\Gamma_Z$ lever, is the subject
of the dedicated analysis in preparation.

All results given here are reproducible from the public release with the archived
analysis code (Appendix~\ref{app:repro}).

\begin{acknowledgments}
The study was wholly conceived and directed at all stages by the author with analysis implementation, execution, and review assisted by the AI systems
Claude (Fable~5 and Fable~5.1, Anthropic) and GPT-6 Astra (OpenAI). The full analysis
and review record is archived and available on request. The author reviewed and verified the analysis products and code and outputs, and takes sole responsibility for the scientific content of the paper. All figures were
produced from analysis of the the public data. We are indebted to the
authors of Ref.~\cite{Cheng:2026sld} for the recovery and public release of the SLD
data and documentation --- this measurement is a direct return on that investment.
This is not an SLD Collaboration result.
\end{acknowledgments}

\appendix

% ======================================================================
\section{Operator matching details}
\label{app:matching}

We work in the minimal SME with Colladay--Kosteleck\'y
conventions~\cite{Colladay:1998fq}: the electron CPT-even coefficients enter per Weyl
chirality as $c_{L,R} = c \pm d$; the post-EWSB $Z$-sector insertions are those of
Ref.~\cite{Fu:2016fmf}, Eq.~(21), with $k_B = -\tan^{2}\theta_W\,k_W$ imposed by the
photon-sector constraints. All matching factors below were obtained by explicit
$\gamma$-matrix trace algebra (computer-verified, archived with the analysis code),
assembling $\sigma_h(q^{2})$ in the improved Born approximation with all
$\mathcal{O}(X)$ ingredients and expanding the ratio $\alr = (\sigma_L-\sigma_R)/
(\sigma_L+\sigma_R)$ to first order.

\emph{Fermion-sector assembly.} For beam helicity $h$, the electron-side chiral tensor
is $E_h^{\nu\nu'} = {\rm Tr}[\tilde{\slashed p}_2\Lambda^{\nu}P_h\tilde{\slashed p}_1
\Lambda^{\nu'}]$ with modified spinor momenta $\tilde p = p + c_h\!\cdot\!p$ and
minimal-coupling vertex $\Lambda^{\nu} = \gamma^{\nu} + c_h^{\mu\nu}\gamma_{\mu}$; the
positron partnering an $e^{-}_{L}$ is a quantum of the same left-chiral field and
carries $c_L$ as well. The bend-magnet lattice fixes the beam \emph{3-momenta}, so the
modified dispersion $(p + c_h\!\cdot\!p)^{2}=0$ shifts the beam energies and hence the
$q^{2}$ presented to the resonance: $\delta q^2/q^2 = -2(c_h^{00} +
c_h^{jk}\hat b_j\hat b_k)$; the Breit--Wigner and interference functions are expanded
about the running point so every slope term carries its exact coefficient. The
final-state quarks and the phase space are conventional (final-state coefficients
cancel in the ratio). The flux normalization was computed in two external-leg
conventions; their difference lands exclusively in the isotropic $c^{00}$ coefficient,
which does not modulate --- every number quoted in the paper is independent of that
choice.
The relative-group-velocity factor of the covariant cross-section
definition~\cite{Colladay:1998fq} \emph{does} enter the observable; an earlier version
of this appendix argued otherwise, and we record the correct accounting. For two
interpenetrating streams the reaction rate per unit volume is $w = n_1 n_2 K_h$ with
$K_h = \int|\mathcal{M}_h|^{2}\,d\Pi/(4E_1E_2)$ --- no velocity appears --- and the
per-crossing yield is $Y_h = \int dt\,d^{3}x\,n_1n_2K_h = N_1N_2\,O_\perp\,K_h/(v_1+v_2)$:
the overlap integral of two counter-propagating bunches gives exactly $1/(v_1+v_2)$
(the Jacobian of $(z,t)\to(z-v_1t,\,z+v_2t)$; verified numerically for unequal
Gaussian bunches), and that factor is the $1/v_{\rm rel}$ of the cross section
$\sigma_h = K_h/(v_1+v_2)$, not the $(v_1+v_2)$ of the luminosity definition. The
earlier argument cancelled the closing velocity inside the luminosity and then also
discarded it from the cross section. For the traceless spatial example
$c_L = {\rm diag}(-d/2,-d/2,d)$ the modified dispersion relation gives the group
velocity $1-d$ exactly for the left-chiral field, and the positron partnering an
$e^-_L$ carries the same coefficient, so the two channels close at $2(1\mp d)$; with
the vertex factors $(1\pm d/2)^{2}$ this gives $Y_L\propto(1+A)(1+d/2)^{2}/(1-d)$,
$Y_R\propto(1-A)(1-d/2)^{2}/(1+d)$, and $\partial\alr/\partial d = 2(1-A^{2})$:
the vertex and the overlap each contribute $1-A^{2}$ (explicit traces of the
electron tensor above give $0.9768$ without and $1.9536$ with the overlap factor at
the Born $A=0.1523$). The velocity factor is present in both machine conventions; only
the dispersion channel below depends on what the machine holds fixed. We note three effects of this accounting in particular: The factor $1/(v_1+v_2)$ is
kinematic and enters the per-crossing yield of every process at the crossing, but
its chiral content is process-specific: the annihilation channel selects the positron
of the electron's own Weyl field, whereas the small-angle Bhabha rates from which SLD
measured its luminosity asymmetry $A_{\mathcal L}$ accept both positron chiralities,
so a luminosity monitor retains only part of the annihilation channel's
closing-velocity response. The factor 2 is therefore specific to the raw counting
asymmetry used here; a monitor-normalized asymmetry would carry a different
$d_e$ response (its overlap contribution is reduced under these weights, and the
full process response is not computed here) and the fixed-energy scenario values
above are observable-specific. A $d_e$-induced closing-velocity modulation has
exactly the signature of the helicity-correlated luminosity modulation $\ell(t)$ of
Sec.~\ref{sec:systematics}, so in the raw asymmetry the two are not separable. And
SLD's DC luminosity asymmetry, $A_{\mathcal L} = (-1.3\pm0.7)\times10^{-4}$,
correspondingly constrains a combination of the exposure-averaged (static, $m=0$)
velocity term $d_e^{\hat b\hat b}$ and instrumental asymmetries, translatable into
the SME coefficient only with the monitor processes' own response --- a DC question
outside this paper's modulation analysis.

\emph{Numerical inputs} (the matching computation, the finite-response scans, and the
azimuthal pilot share them): $M_Z = 91.1876$~GeV, $\Gamma_Z = 2.4952$~GeV,
$\sin^{2}\theta_{\rm eff} = 0.23098$, $\sqrt s = 91.238$~GeV (the luminosity-weighted SLC
point), five massless quark flavours with colour factor 3, $Q_e=-1$, improved-Born
couplings $x_{L} = -\tfrac12 + \sin^{2}\theta_{\rm eff}$, $x_{R} = \sin^{2}\theta_{\rm eff}$ with the $Z$ normalization
$1/[\sin^{2}\theta_{\rm eff}(1-\sin^{2}\theta_{\rm eff})]$, constant width, giving
$A_e = 0.1513$ and the Born $\alr = 0.1523$; the photon-exchange term is single-counted
in the archived matching code (a chirality-summed photon term changes the factors
quoted here at the $0.05$--$0.1\%$ level).

\emph{Internal validations.} The assembly reproduces the improved-Born
$\alr(\sqrt{s})$ (value, $A_e$, and slope $+0.022$~GeV$^{-1}$) at zeroth order; the
transverse $c_h$ coefficients emerge as exactly $\mp(1-A^{2})/2$ in the
$Z$-exchange-only assembly (the chiral
enhancement, derived rather than assumed; this is the vertex-and-spinor part, to which
the per-crossing closing-velocity factor above adds an equal amount); the
chirality-even longitudinal coefficient
equals $-2(d\alr/d\sqrt s)(\sqrt s/2)$ to $0.2\%$, confirming that $c_e$ survives only
through the energy-shift channel, as required by the coordinate-redefinition argument;
and the $\gamma$--$Z$ mixing shift is helicity-independent component-by-component,
with $\partial\alr/\partial(\delta g) = 2[(x_L-x_R) - \bar A(x_L+x_R)]/(x_L^2+x_R^2)
= -7.86$ matching the analytic form.

\emph{Results.} With $\hat q$ purely timelike, the mixing insertion contracts to
$K^{\lambda\nu} = s\,\kw^{T\lambda T\nu}$; the transverse electron currents give
$\delta\alr^{\rm mix}(t) = -1.815\,\kw^{T\hat bT\hat b}(t)$, with all
longitudinal and off-diagonal contractions vanishing. (Without imposing $k_F\simeq0$
the probed combination is $[\cos^{2}\theta_W\,(k_W - k_B)]^{TJTK}$, which the
photon-sector constraints reduce to $\kw^{TJTK}$ to far better than our precision.
The partner diagram with the mixing on the quark side shifts the $f$ couplings
identically for both helicities and survives only at the interference/photon fraction
of the cross section, $\sim10^{-3}$ of the electron-side term.) The $ZZ$-kinetic
insertion acts through the $\kpp$ channel as $\kpp^{\rm eff} =
-2(1-\tan^{2}\theta_W)(s/M_Z^{2})\kw^{T\lambda T\nu}$, contributing
$\delta\alr^{ZZ}(t) = +0.699\,\kw^{T\hat bT\hat b}(t)$; the total, $-1.116$, is
stable to $10^{-3}$ across the $\pm29$~MeV energy-scale band, because the surviving
$\kpp$-channel structure ${\rm Re}\,D^{2}/|D|^{2}\to-1$ is detuning-independent for
$\Delta^{2}\ll M_Z^{2}\Gamma_Z^{2}$. These propagator-channel factors are Born-level;
the ISR-dressed derivative of the ratio observable carries the dressing stated
in Sec.~\ref{sec:matching} ($-8.5\%$ total for $\kpp$, $-8.6\%$ for the $ZZ$-kinetic
part, the covariance term dominating),
computed with the $\mathcal{O}(\alpha)$-exponentiated radiator. The $\kpp$ insertion
between conserved currents cancels in the $|Z|^{2}$ term (a chirality-blind rescale)
and survives through the $\gamma$--$Z$ interference: $\delta\alr(t) =
-0.499\,\kpp^{\hat b\hat b}(t)$. For $d_e$,
$\delta\alr = d^{jk}[-(1-\bar A^{2})\delta^{jk} + \beta\,\hat b^{j}\hat b^{k}]$; the
$\delta^{jk}$ term is rotation-invariant, so the modulation is
$\beta\,d^{\hat b\hat b}(t)$. The trace computation gives the vertex part
($\beta_{\rm vtx} = 1-A^{2}$ exactly, $=0.9768$ at the Born $A = 0.1523$; the archived
running-point value $0.9775$ is a 10~MeV finite-difference diagnostic that converges to
it as the step shrinks and is retained in the quoted $\beta_0$), and the closing-velocity
factor of the per-crossing yield
doubles it: $\beta_0 = 2\beta_{\rm vtx} = 1.955$. The dispersion channel
depends on
what the machine holds fixed: the SLC arcs and energy spectrometers define the
delivered beams through magnetic rigidity, i.e.\ momentum, and in this fixed-momentum
convention the dispersion term adds $-\ecm[\beta_{\rm vtx}\,\sigma'/\sigma -
\bar A\,d\alr/d\ecm] = -3.9\pm2.8$ with SLD's dressed slope $\ecm\sigma'/\sigma =
+4.3\pm2.9$ (positive: the observed cross section rises at $91.237$~GeV,
Sec.~\ref{sec:matching}), giving $\beta = -1.96\pm2.83$; in a fixed-energy convention
the dispersion channel is absent and $\beta = \beta_0$. The profile likelihood over
$(\rho_{d},\varphi,\beta)$ with $\beta$ as a Gaussian constraint (the
energy-scale/lineshape knowledge) returns no limit: decoupling ($\beta\to0$) costs
$\Delta\chi^{2}=(\beta/\sigma_\beta)^{2}=0.48$, below the one-sided 90\% ($1.64$) and
95\% ($2.71$) thresholds, for both harmonics (the second-harmonic profile uses the
response-corrected binned $(C_2,S_2)$ as input, Sec.~\ref{sec:method}). The
fixed-energy scenario envelopes,
$0.0086/(0.982\,\beta_0) = 4.5\times10^{-3}$ and $0.0107/(0.406\,\beta_0) =
1.35\times10^{-2}$, are quoted in Sec.~\ref{sec:matching} as conditional scenarios,
not limits. The $k_W$ and $\kpp$ linear factors are dispersion-free and
convention-immune; their finite-response treatment is in Sec.~\ref{sec:matching}.
Boost-mixed components ($c^{0j}$, $d^{0j}$)
cancel identically between the two symmetric beams (a symmetric collider is blind
to them at leading order) and, with the CPT-odd gauge terms $k_{1,2}$ set to zero
(Sec.~\ref{sec:matching}), no operator of the minimal SME populates the rank-one
channel at $\mathcal{O}(1)$: $b_e$ is suppressed by $m_e/E_b\simeq1.1\times10^{-5}$,
and constant vector
backgrounds (including the $Z$ vacuum value induced by $k_{\phi}$) decouple from
chirality-diagonal processes by phase redefinition. The complete
derivation, with the
coefficient-by-coefficient output of the trace computation and the full uncertainty
ledger, is archived with the analysis (Appendix~\ref{app:repro}).

\section{Reproducibility}
\label{app:repro}

The data are the public Zenodo records (v1: 10.5281/zenodo.19925960; v2:
10.5281/zenodo.21199778). The event samples used
here are defined by: the release's \texttt{alr\_2000} hadronic and 2001 leptonic
selection presets; validity requirements $|P_e|>0.1$ and
$757382400 < \texttt{evttime} < 915148800$ (Unix UTC); run blocks $=$ mini-DST file
groups. The joint likelihood is Eq.~(\ref{eq:model}) with the binomial per-event
term; all fits, binnings, and phase conventions are fixed in the archived analysis
code (Sec.~\ref{sec:dataset}) at
\href{https://doi.org/10.5281/zenodo.23599014}{doi:10.5281/zenodo.23599014}.

\bibliography{refs}

\begin{thebibliography}{47}%
\makeatletter
\providecommand \@ifxundefined [1]{%
 \@ifx{#1\undefined}
}%
\providecommand \@ifnum [1]{%
 \ifnum #1\expandafter \@firstoftwo
 \else \expandafter \@secondoftwo
 \fi
}%
\providecommand \@ifx [1]{%
 \ifx #1\expandafter \@firstoftwo
 \else \expandafter \@secondoftwo
 \fi
}%
\providecommand \natexlab [1]{#1}%
\providecommand \enquote  [1]{``#1''}%
\providecommand \bibnamefont  [1]{#1}%
\providecommand \bibfnamefont [1]{#1}%
\providecommand \citenamefont [1]{#1}%
\providecommand \href@noop [0]{\@secondoftwo}%
\providecommand \href [0]{\begingroup \@sanitize@url \@href}%
\providecommand \@href[1]{\@@startlink{#1}\@@href}%
\providecommand \@@href[1]{\endgroup#1\@@endlink}%
\providecommand \@sanitize@url [0]{\catcode `\\12\catcode `\$12\catcode
  `\&12\catcode `\#12\catcode `\^12\catcode `\_12\catcode `\%12\relax}%
\providecommand \@@startlink[1]{}%
\providecommand \@@endlink[0]{}%
\providecommand \url  [0]{\begingroup\@sanitize@url \@url }%
\providecommand \@url [1]{\endgroup\@href {#1}{\urlprefix }}%
\providecommand \urlprefix  [0]{URL }%
\providecommand \Eprint [0]{\href }%
\providecommand \doibase [0]{https://doi.org/}%
\providecommand \selectlanguage [0]{\@gobble}%
\providecommand \bibinfo  [0]{\@secondoftwo}%
\providecommand \bibfield  [0]{\@secondoftwo}%
\providecommand \translation [1]{[#1]}%
\providecommand \BibitemOpen [0]{}%
\providecommand \bibitemStop [0]{}%
\providecommand \bibitemNoStop [0]{.\EOS\space}%
\providecommand \EOS [0]{\spacefactor3000\relax}%
\providecommand \BibitemShut  [1]{\csname bibitem#1\endcsname}%
\let\auto@bib@innerbib\@empty
%</preamble>
\bibitem [{\citenamefont {Colladay}\ and\ \citenamefont
  {Kosteleck\'y}(1997)}]{Colladay:1996iz}%
  \BibitemOpen
  \bibfield  {author} {\bibinfo {author} {\bibfnamefont {D.}~\bibnamefont
  {Colladay}}\ and\ \bibinfo {author} {\bibfnamefont {V.~A.}\ \bibnamefont
  {Kosteleck\'y}},\ }\bibfield  {title} {\bibinfo {title} {{CPT violation and
  the standard model}},\ }\href@noop {} {\bibfield  {journal} {\bibinfo
  {journal} {Phys. Rev. D}\ }\textbf {\bibinfo {volume} {55}},\ \bibinfo
  {pages} {6760} (\bibinfo {year} {1997})},\ \Eprint
  {https://arxiv.org/abs/hep-ph/9703464} {arXiv:hep-ph/9703464} \BibitemShut
  {NoStop}%
\bibitem [{\citenamefont {Colladay}\ and\ \citenamefont
  {Kosteleck\'y}(1998)}]{Colladay:1998fq}%
  \BibitemOpen
  \bibfield  {author} {\bibinfo {author} {\bibfnamefont {D.}~\bibnamefont
  {Colladay}}\ and\ \bibinfo {author} {\bibfnamefont {V.~A.}\ \bibnamefont
  {Kosteleck\'y}},\ }\bibfield  {title} {\bibinfo {title} {{Lorentz-violating
  extension of the standard model}},\ }\href@noop {} {\bibfield  {journal}
  {\bibinfo  {journal} {Phys. Rev. D}\ }\textbf {\bibinfo {volume} {58}},\
  \bibinfo {pages} {116002} (\bibinfo {year} {1998})},\ \Eprint
  {https://arxiv.org/abs/hep-ph/9809521} {arXiv:hep-ph/9809521} \BibitemShut
  {NoStop}%
\bibitem [{\citenamefont {Kosteleck\'y}\ and\ \citenamefont
  {Russell}(2011)}]{Kostelecky:2008ts}%
  \BibitemOpen
  \bibfield  {author} {\bibinfo {author} {\bibfnamefont {V.~A.}\ \bibnamefont
  {Kosteleck\'y}}\ and\ \bibinfo {author} {\bibfnamefont {N.}~\bibnamefont
  {Russell}},\ }\bibfield  {title} {\bibinfo {title} {{Data Tables for Lorentz
  and CPT Violation}},\ }\href@noop {} {\bibfield  {journal} {\bibinfo
  {journal} {Rev. Mod. Phys.}\ }\textbf {\bibinfo {volume} {83}},\ \bibinfo
  {pages} {11} (\bibinfo {year} {2011})},\ \bibinfo {note} {annually updated
  edition; Table D46 (electroweak sector) as fetched 2026-07-07},\ \Eprint
  {https://arxiv.org/abs/0801.0287} {arXiv:0801.0287} \BibitemShut {NoStop}%
\bibitem [{\citenamefont {Kosteleck\'y}\ and\ \citenamefont
  {Mewes}(2002)}]{Kostelecky:2002hh}%
  \BibitemOpen
  \bibfield  {author} {\bibinfo {author} {\bibfnamefont {V.~A.}\ \bibnamefont
  {Kosteleck\'y}}\ and\ \bibinfo {author} {\bibfnamefont {M.}~\bibnamefont
  {Mewes}},\ }\bibfield  {title} {\bibinfo {title} {{Signals for Lorentz
  violation in electrodynamics}},\ }\href@noop {} {\bibfield  {journal}
  {\bibinfo  {journal} {Phys. Rev. D}\ }\textbf {\bibinfo {volume} {66}},\
  \bibinfo {pages} {056005} (\bibinfo {year} {2002})},\ \bibinfo {note}
  {sun-centered celestial equatorial frame conventions},\ \Eprint
  {https://arxiv.org/abs/hep-ph/0205211} {arXiv:hep-ph/0205211} \BibitemShut
  {NoStop}%
\bibitem [{\citenamefont {Abazov}\ \emph {et~al.}(2012)\citenamefont {Abazov}
  \emph {et~al.}}]{D0:2012rbu}%
  \BibitemOpen
  \bibfield  {author} {\bibinfo {author} {\bibfnamefont {V.~M.}\ \bibnamefont
  {Abazov}} \emph {et~al.} (\bibinfo {collaboration} {D0}),\ }\bibfield
  {title} {\bibinfo {title} {{Search for violation of Lorentz invariance in top
  quark pair production and decay}},\ }\href@noop {} {\bibfield  {journal}
  {\bibinfo  {journal} {Phys. Rev. Lett.}\ }\textbf {\bibinfo {volume} {108}},\
  \bibinfo {pages} {261603} (\bibinfo {year} {2012})},\ \Eprint
  {https://arxiv.org/abs/1203.6106} {arXiv:1203.6106} \BibitemShut {NoStop}%
\bibitem [{\citenamefont {Hayrapetyan}\ \emph {et~al.}(2024)\citenamefont
  {Hayrapetyan} \emph {et~al.}}]{CMS:2024tt}%
  \BibitemOpen
  \bibfield  {author} {\bibinfo {author} {\bibfnamefont {A.}~\bibnamefont
  {Hayrapetyan}} \emph {et~al.} (\bibinfo {collaboration} {CMS}),\ }\bibfield
  {title} {\bibinfo {title} {{Searches for violation of Lorentz invariance in
  top quark pair production using dilepton events in 13 TeV proton-proton
  collisions}},\ }\href {https://doi.org/10.1016/j.physletb.2024.138979}
  {\bibfield  {journal} {\bibinfo  {journal} {Phys. Lett. B}\ }\textbf
  {\bibinfo {volume} {857}},\ \bibinfo {pages} {138979} (\bibinfo {year}
  {2024})},\ \Eprint {https://arxiv.org/abs/2405.14757} {arXiv:2405.14757}
  \BibitemShut {NoStop}%
\bibitem [{\citenamefont {Abt}\ \emph {et~al.}(2023)\citenamefont {Abt} \emph
  {et~al.}}]{ZEUS:2022pbe}%
  \BibitemOpen
  \bibfield  {author} {\bibinfo {author} {\bibfnamefont {I.}~\bibnamefont
  {Abt}} \emph {et~al.} (\bibinfo {collaboration} {ZEUS}),\ }\bibfield  {title}
  {\bibinfo {title} {{Search for effective Lorentz and CPT violation using ZEUS
  data}},\ }\href {https://doi.org/10.1103/PhysRevD.107.092008} {\bibfield
  {journal} {\bibinfo  {journal} {Phys. Rev. D}\ }\textbf {\bibinfo {volume}
  {107}},\ \bibinfo {pages} {092008} (\bibinfo {year} {2023})},\ \Eprint
  {https://arxiv.org/abs/2212.12750} {arXiv:2212.12750} \BibitemShut {NoStop}%
\bibitem [{\citenamefont {Kosteleck\'y}\ \emph {et~al.}(2017)\citenamefont
  {Kosteleck\'y}, \citenamefont {Lunghi},\ and\ \citenamefont
  {Vieira}}]{Kostelecky:2016pyx}%
  \BibitemOpen
  \bibfield  {author} {\bibinfo {author} {\bibfnamefont {V.~A.}\ \bibnamefont
  {Kosteleck\'y}}, \bibinfo {author} {\bibfnamefont {E.}~\bibnamefont
  {Lunghi}},\ and\ \bibinfo {author} {\bibfnamefont {A.~R.}\ \bibnamefont
  {Vieira}},\ }\bibfield  {title} {\bibinfo {title} {{Lorentz violation and
  deep inelastic scattering}},\ }\href@noop {} {\bibfield  {journal} {\bibinfo
  {journal} {Phys. Lett. B}\ }\textbf {\bibinfo {volume} {769}},\ \bibinfo
  {pages} {272} (\bibinfo {year} {2017})},\ \Eprint
  {https://arxiv.org/abs/1610.08755} {arXiv:1610.08755} \BibitemShut {NoStop}%
\bibitem [{\citenamefont {Lunghi}\ and\ \citenamefont
  {Sherrill}(2018)}]{Lunghi:2018uwj}%
  \BibitemOpen
  \bibfield  {author} {\bibinfo {author} {\bibfnamefont {E.}~\bibnamefont
  {Lunghi}}\ and\ \bibinfo {author} {\bibfnamefont {N.}~\bibnamefont
  {Sherrill}},\ }\bibfield  {title} {\bibinfo {title} {{Lorentz violation and
  the electron-ion collider}},\ }\href
  {https://doi.org/10.1103/PhysRevD.98.115018} {\bibfield  {journal} {\bibinfo
  {journal} {Phys. Rev. D}\ }\textbf {\bibinfo {volume} {98}},\ \bibinfo
  {pages} {115018} (\bibinfo {year} {2018})},\ \Eprint
  {https://arxiv.org/abs/1805.11684} {arXiv:1805.11684} \BibitemShut {NoStop}%
\bibitem [{\citenamefont {Kosteleck\'y}\ and\ \citenamefont
  {Li}(2019)}]{Kostelecky:2018yfa}%
  \BibitemOpen
  \bibfield  {author} {\bibinfo {author} {\bibfnamefont {V.~A.}\ \bibnamefont
  {Kosteleck\'y}}\ and\ \bibinfo {author} {\bibfnamefont {Z.}~\bibnamefont
  {Li}},\ }\bibfield  {title} {\bibinfo {title} {{Gauge field theories with
  Lorentz-violating operators of arbitrary dimension}},\ }\href@noop {}
  {\bibfield  {journal} {\bibinfo  {journal} {Phys. Rev. D}\ }\textbf {\bibinfo
  {volume} {99}},\ \bibinfo {pages} {056016} (\bibinfo {year} {2019})},\
  \Eprint {https://arxiv.org/abs/1812.11672} {arXiv:1812.11672} \BibitemShut
  {NoStop}%
\bibitem [{\citenamefont {Lunghi}\ \emph {et~al.}(2021)\citenamefont {Lunghi},
  \citenamefont {Sherrill}, \citenamefont {Szczepaniak},\ and\ \citenamefont
  {Vieira}}]{Lunghi:2021jhep}%
  \BibitemOpen
  \bibfield  {author} {\bibinfo {author} {\bibfnamefont {E.}~\bibnamefont
  {Lunghi}}, \bibinfo {author} {\bibfnamefont {N.}~\bibnamefont {Sherrill}},
  \bibinfo {author} {\bibfnamefont {A.}~\bibnamefont {Szczepaniak}},\ and\
  \bibinfo {author} {\bibfnamefont {A.}~\bibnamefont {Vieira}},\ }\bibfield
  {title} {\bibinfo {title} {{Quark-sector Lorentz violation in $Z$-boson
  production}},\ }\href@noop {} {\bibfield  {journal} {\bibinfo  {journal}
  {JHEP}\ }\textbf {\bibinfo {volume} {04}},\ \bibinfo {pages} {228}},\
  \bibinfo {note} {erratum: JHEP \textbf{07} (2024) 167},\ \Eprint
  {https://arxiv.org/abs/2011.02632} {arXiv:2011.02632} \BibitemShut {NoStop}%
\bibitem [{\citenamefont {Link}\ \emph {et~al.}(2003)\citenamefont {Link} \emph
  {et~al.}}]{FOCUS:2003cpt}%
  \BibitemOpen
  \bibfield  {author} {\bibinfo {author} {\bibfnamefont {J.~M.}\ \bibnamefont
  {Link}} \emph {et~al.} (\bibinfo {collaboration} {FOCUS}),\ }\bibfield
  {title} {\bibinfo {title} {{Charm System Tests of CPT and Lorentz Invariance
  with FOCUS}},\ }\href {https://doi.org/10.1016/S0370-2693(03)00103-5}
  {\bibfield  {journal} {\bibinfo  {journal} {Phys. Lett. B}\ }\textbf
  {\bibinfo {volume} {556}},\ \bibinfo {pages} {7} (\bibinfo {year} {2003})},\
  \Eprint {https://arxiv.org/abs/hep-ex/0208034} {arXiv:hep-ex/0208034}
  \BibitemShut {NoStop}%
\bibitem [{\citenamefont {Aubert}\ \emph {et~al.}(2008)\citenamefont {Aubert}
  \emph {et~al.}}]{BaBar:2008cpt}%
  \BibitemOpen
  \bibfield  {author} {\bibinfo {author} {\bibfnamefont {B.}~\bibnamefont
  {Aubert}} \emph {et~al.} (\bibinfo {collaboration} {BaBar}),\ }\bibfield
  {title} {\bibinfo {title} {{Search for CPT and Lorentz violation in
  $B^0$--$\bar{B}^0$ oscillations with dilepton events}},\ }\href
  {https://doi.org/10.1103/PhysRevLett.100.131802} {\bibfield  {journal}
  {\bibinfo  {journal} {Phys. Rev. Lett.}\ }\textbf {\bibinfo {volume} {100}},\
  \bibinfo {pages} {131802} (\bibinfo {year} {2008})},\ \Eprint
  {https://arxiv.org/abs/0711.2713} {arXiv:0711.2713} \BibitemShut {NoStop}%
\bibitem [{\citenamefont {Abazov}\ \emph {et~al.}(2015)\citenamefont {Abazov}
  \emph {et~al.}}]{D0:2015cpt}%
  \BibitemOpen
  \bibfield  {author} {\bibinfo {author} {\bibfnamefont {V.~M.}\ \bibnamefont
  {Abazov}} \emph {et~al.} (\bibinfo {collaboration} {D0}),\ }\bibfield
  {title} {\bibinfo {title} {{Search for Violation of $CPT$ and Lorentz
  Invariance in ${B_s^0}$ Meson Oscillations}},\ }\href
  {https://doi.org/10.1103/PhysRevLett.115.161601} {\bibfield  {journal}
  {\bibinfo  {journal} {Phys. Rev. Lett.}\ }\textbf {\bibinfo {volume} {115}},\
  \bibinfo {pages} {161601} (\bibinfo {year} {2015})},\ \bibinfo {note}
  {addendum: Phys. Rev. Lett. \textbf{116}, 019901 (2016)},\ \Eprint
  {https://arxiv.org/abs/1506.04123} {arXiv:1506.04123} \BibitemShut {NoStop}%
\bibitem [{\citenamefont {Aaij}\ \emph {et~al.}(2016)\citenamefont {Aaij} \emph
  {et~al.}}]{LHCb:2016cpt}%
  \BibitemOpen
  \bibfield  {author} {\bibinfo {author} {\bibfnamefont {R.}~\bibnamefont
  {Aaij}} \emph {et~al.} (\bibinfo {collaboration} {LHCb}),\ }\bibfield
  {title} {\bibinfo {title} {{Search for violations of Lorentz invariance and
  $CPT$ symmetry in $B^0_{(s)}$ mixing}},\ }\href
  {https://doi.org/10.1103/PhysRevLett.116.241601} {\bibfield  {journal}
  {\bibinfo  {journal} {Phys. Rev. Lett.}\ }\textbf {\bibinfo {volume} {116}},\
  \bibinfo {pages} {241601} (\bibinfo {year} {2016})},\ \Eprint
  {https://arxiv.org/abs/1603.04804} {arXiv:1603.04804} \BibitemShut {NoStop}%
\bibitem [{\citenamefont {Babusci}\ \emph {et~al.}(2014)\citenamefont {Babusci}
  \emph {et~al.}}]{KLOE:2014cpt}%
  \BibitemOpen
  \bibfield  {author} {\bibinfo {author} {\bibfnamefont {D.}~\bibnamefont
  {Babusci}} \emph {et~al.} (\bibinfo {collaboration} {KLOE-2}),\ }\bibfield
  {title} {\bibinfo {title} {{Test of CPT and Lorentz symmetry in entangled
  neutral kaons with the KLOE experiment}},\ }\href
  {https://doi.org/10.1016/j.physletb.2014.01.026} {\bibfield  {journal}
  {\bibinfo  {journal} {Phys. Lett. B}\ }\textbf {\bibinfo {volume} {730}},\
  \bibinfo {pages} {89} (\bibinfo {year} {2014})},\ \Eprint
  {https://arxiv.org/abs/1312.6818} {arXiv:1312.6818} \BibitemShut {NoStop}%
\bibitem [{\citenamefont {M{\"u}ller}\ \emph {et~al.}(2013)\citenamefont
  {M{\"u}ller}, \citenamefont {Dijck}, \citenamefont {Bekker} \emph
  {et~al.}}]{Muller:2013wck}%
  \BibitemOpen
  \bibfield  {author} {\bibinfo {author} {\bibfnamefont {S.~E.}\ \bibnamefont
  {M{\"u}ller}}, \bibinfo {author} {\bibfnamefont {E.~A.}\ \bibnamefont
  {Dijck}}, \bibinfo {author} {\bibfnamefont {H.}~\bibnamefont {Bekker}}, \emph
  {et~al.},\ }\bibfield  {title} {\bibinfo {title} {{First Test of Lorentz
  Invariance in the Weak Decay of Polarized Nuclei}},\ }\href
  {https://doi.org/10.1103/PhysRevD.88.071901} {\bibfield  {journal} {\bibinfo
  {journal} {Phys. Rev. D}\ }\textbf {\bibinfo {volume} {88}},\ \bibinfo
  {pages} {071901} (\bibinfo {year} {2013})},\ \Eprint
  {https://arxiv.org/abs/1310.3206} {arXiv:1310.3206 [nucl-ex]} \BibitemShut
  {NoStop}%
\bibitem [{\citenamefont {Sytema}\ \emph {et~al.}(2016)\citenamefont {Sytema},
  \citenamefont {van~den Berg}, \citenamefont {B{\"o}ll} \emph
  {et~al.}}]{Sytema:2016xiv}%
  \BibitemOpen
  \bibfield  {author} {\bibinfo {author} {\bibfnamefont {A.}~\bibnamefont
  {Sytema}}, \bibinfo {author} {\bibfnamefont {J.~E.}\ \bibnamefont {van~den
  Berg}}, \bibinfo {author} {\bibfnamefont {O.}~\bibnamefont {B{\"o}ll}}, \emph
  {et~al.},\ }\bibfield  {title} {\bibinfo {title} {{Test of Lorentz invariance
  in $\beta$ decay of polarized $^{20}$Na}},\ }\href
  {https://doi.org/10.1103/PhysRevC.94.025503} {\bibfield  {journal} {\bibinfo
  {journal} {Phys. Rev. C}\ }\textbf {\bibinfo {volume} {94}},\ \bibinfo
  {pages} {025503} (\bibinfo {year} {2016})},\ \Eprint
  {https://arxiv.org/abs/1608.03870} {arXiv:1608.03870 [nucl-ex]} \BibitemShut
  {NoStop}%
\bibitem [{\citenamefont {Noordmans}\ \emph
  {et~al.}(2013{\natexlab{a}})\citenamefont {Noordmans}, \citenamefont
  {Wilschut},\ and\ \citenamefont
  {Timmermans}}]{NoordmansWilschutTimmermans:2013}%
  \BibitemOpen
  \bibfield  {author} {\bibinfo {author} {\bibfnamefont {J.~P.}\ \bibnamefont
  {Noordmans}}, \bibinfo {author} {\bibfnamefont {H.~W.}\ \bibnamefont
  {Wilschut}},\ and\ \bibinfo {author} {\bibfnamefont {R.~G.~E.}\ \bibnamefont
  {Timmermans}},\ }\bibfield  {title} {\bibinfo {title} {{Limits on Lorentz
  violation from forbidden $\beta$ decays}},\ }\href@noop {} {\bibfield
  {journal} {\bibinfo  {journal} {Phys. Rev. Lett.}\ }\textbf {\bibinfo
  {volume} {111}},\ \bibinfo {pages} {171601} (\bibinfo {year}
  {2013}{\natexlab{a}})},\ \Eprint {https://arxiv.org/abs/1308.5570}
  {arXiv:1308.5570} \BibitemShut {NoStop}%
\bibitem [{\citenamefont {Noordmans}\ \emph
  {et~al.}(2013{\natexlab{b}})\citenamefont {Noordmans}, \citenamefont
  {Wilschut},\ and\ \citenamefont {Timmermans}}]{Noordmans:2013oia}%
  \BibitemOpen
  \bibfield  {author} {\bibinfo {author} {\bibfnamefont {J.~P.}\ \bibnamefont
  {Noordmans}}, \bibinfo {author} {\bibfnamefont {H.~W.}\ \bibnamefont
  {Wilschut}},\ and\ \bibinfo {author} {\bibfnamefont {R.~G.~E.}\ \bibnamefont
  {Timmermans}},\ }\bibfield  {title} {\bibinfo {title} {{Lorentz violation in
  neutron decay and allowed nuclear $\beta$ decay}},\ }\href@noop {} {\bibfield
   {journal} {\bibinfo  {journal} {Phys. Rev. C}\ }\textbf {\bibinfo {volume}
  {87}},\ \bibinfo {pages} {055502} (\bibinfo {year} {2013}{\natexlab{b}})},\
  \Eprint {https://arxiv.org/abs/1302.2730} {arXiv:1302.2730} \BibitemShut
  {NoStop}%
\bibitem [{\citenamefont {Fu}\ and\ \citenamefont
  {Lehnert}(2016)}]{Fu:2016fmf}%
  \BibitemOpen
  \bibfield  {author} {\bibinfo {author} {\bibfnamefont {H.}~\bibnamefont
  {Fu}}\ and\ \bibinfo {author} {\bibfnamefont {R.}~\bibnamefont {Lehnert}},\
  }\bibfield  {title} {\bibinfo {title} {{M{\o}ller scattering and
  Lorentz-violating Z bosons}},\ }\href
  {https://doi.org/10.1016/j.physletb.2016.08.063} {\bibfield  {journal}
  {\bibinfo  {journal} {Phys. Lett. B}\ }\textbf {\bibinfo {volume} {762}},\
  \bibinfo {pages} {33} (\bibinfo {year} {2016})}\BibitemShut {NoStop}%
\bibitem [{\citenamefont {Altschul}(2007)}]{Altschul:2007kr}%
  \BibitemOpen
  \bibfield  {author} {\bibinfo {author} {\bibfnamefont {B.}~\bibnamefont
  {Altschul}},\ }\bibfield  {title} {\bibinfo {title} {{Astrophysical limits on
  Lorentz violation for all charged species}},\ }\href
  {https://doi.org/10.1016/j.astropartphys.2007.08.003} {\bibfield  {journal}
  {\bibinfo  {journal} {Astropart. Phys.}\ }\textbf {\bibinfo {volume} {28}},\
  \bibinfo {pages} {380} (\bibinfo {year} {2007})},\ \Eprint
  {https://arxiv.org/abs/hep-ph/0610324} {arXiv:hep-ph/0610324} \BibitemShut
  {NoStop}%
\bibitem [{\citenamefont {Iltan}(2005)}]{Iltan:2005zll}%
  \BibitemOpen
  \bibfield  {author} {\bibinfo {author} {\bibfnamefont {E.~O.}\ \bibnamefont
  {Iltan}},\ }\bibfield  {title} {\bibinfo {title} {{The Lorentz and CPT
  violating effects on the $Z \to l^{+} l^{-}$ decay}},\ }\href@noop {}
  {\bibfield  {journal} {\bibinfo  {journal} {Eur. Phys. J. C}\ }\textbf
  {\bibinfo {volume} {40}},\ \bibinfo {pages} {269} (\bibinfo {year} {2005})},\
  \Eprint {https://arxiv.org/abs/hep-ph/0308151} {arXiv:hep-ph/0308151}
  \BibitemShut {NoStop}%
\bibitem [{\citenamefont {Michel}\ and\ \citenamefont
  {Sher}(2019)}]{Michel:2019tti}%
  \BibitemOpen
  \bibfield  {author} {\bibinfo {author} {\bibfnamefont {A.}~\bibnamefont
  {Michel}}\ and\ \bibinfo {author} {\bibfnamefont {M.}~\bibnamefont {Sher}},\
  }\bibfield  {title} {\bibinfo {title} {{Breaking of Lorentz Invariance in
  Electron-Proton Parity Violation}},\ }\href
  {https://doi.org/10.1103/PhysRevD.100.095011} {\bibfield  {journal} {\bibinfo
   {journal} {Phys. Rev. D}\ }\textbf {\bibinfo {volume} {100}},\ \bibinfo
  {pages} {095011} (\bibinfo {year} {2019})},\ \Eprint
  {https://arxiv.org/abs/1909.10627} {arXiv:1909.10627} \BibitemShut {NoStop}%
\bibitem [{\citenamefont {Cheng}\ \emph
  {et~al.}(2026{\natexlab{a}})\citenamefont {Cheng}, \citenamefont {Corrodi},
  \citenamefont {Hobbs}, \citenamefont {Mete},\ and\ \citenamefont
  {Nachman}}]{Cheng:2026sld}%
  \BibitemOpen
  \bibfield  {author} {\bibinfo {author} {\bibfnamefont {C.~L.}\ \bibnamefont
  {Cheng}}, \bibinfo {author} {\bibfnamefont {S.}~\bibnamefont {Corrodi}},
  \bibinfo {author} {\bibfnamefont {T.~J.}\ \bibnamefont {Hobbs}}, \bibinfo
  {author} {\bibfnamefont {A.~S.}\ \bibnamefont {Mete}},\ and\ \bibinfo
  {author} {\bibfnamefont {B.}~\bibnamefont {Nachman}},\ }\bibfield  {title}
  {\bibinfo {title} {{An AI-ready, Polarized Electron-Positron Collision
  Dataset}},\ }\href@noop {} {\  (\bibinfo {year} {2026}{\natexlab{a}})},\
  \bibinfo {note} {dataset: Zenodo,
  \href{https://doi.org/10.5281/zenodo.19925960}{10.5281/zenodo.19925960} (v1);
  \href{https://doi.org/10.5281/zenodo.21199778}{10.5281/zenodo.21199778} (v2,
  full 1996--1998)},\ \Eprint {https://arxiv.org/abs/2606.00224}
  {arXiv:2606.00224 [hep-ex]} \BibitemShut {NoStop}%
\bibitem [{\citenamefont {Cheng}\ \emph
  {et~al.}(2026{\natexlab{b}})\citenamefont {Cheng}, \citenamefont {Nachman},
  \citenamefont {Mete}, \citenamefont {Hobbs}, \citenamefont {Corrodi},
  \citenamefont {Johnson}, \citenamefont {Breidenbach},\ and\ \citenamefont
  {Su}}]{zenodo:v2}%
  \BibitemOpen
  \bibfield  {author} {\bibinfo {author} {\bibfnamefont {C.~L.}\ \bibnamefont
  {Cheng}}, \bibinfo {author} {\bibfnamefont {B.}~\bibnamefont {Nachman}},
  \bibinfo {author} {\bibfnamefont {A.~S.}\ \bibnamefont {Mete}}, \bibinfo
  {author} {\bibfnamefont {T.}~\bibnamefont {Hobbs}}, \bibinfo {author}
  {\bibfnamefont {S.}~\bibnamefont {Corrodi}}, \bibinfo {author} {\bibfnamefont
  {A.}~\bibnamefont {Johnson}}, \bibinfo {author} {\bibfnamefont
  {M.}~\bibnamefont {Breidenbach}},\ and\ \bibinfo {author} {\bibfnamefont
  {D.}~\bibnamefont {Su}},\ }\href {https://doi.org/10.5281/zenodo.21199778}
  {\bibinfo {title} {{SLD reconstructed mini-DSTs from the 1996--1998 SLC runs
  and documentation corpus}}} (\bibinfo {year} {2026}{\natexlab{b}}),\ \bibinfo
  {note} {version 2, published 2026-07-05. Files used:
  \texttt{sld\_minidsts\_parquet\_1996\_1998.zip} (md5
  \texttt{a2df00a6\ldots}), \texttt{sld\_corpus.zip} (md5
  \texttt{2e2bb361\ldots})}\BibitemShut {NoStop}%
\bibitem [{\citenamefont {Abe}\ \emph {et~al.}(2000)\citenamefont {Abe} \emph
  {et~al.}}]{SLD:2000jop}%
  \BibitemOpen
  \bibfield  {author} {\bibinfo {author} {\bibfnamefont {K.}~\bibnamefont
  {Abe}} \emph {et~al.} (\bibinfo {collaboration} {SLD}),\ }\bibfield  {title}
  {\bibinfo {title} {{A high precision measurement of the left-right Z boson
  cross-section asymmetry}},\ }\href@noop {} {\bibfield  {journal} {\bibinfo
  {journal} {Phys. Rev. Lett.}\ }\textbf {\bibinfo {volume} {84}},\ \bibinfo
  {pages} {5945} (\bibinfo {year} {2000})},\ \Eprint
  {https://arxiv.org/abs/hep-ex/0004026} {arXiv:hep-ex/0004026} \BibitemShut
  {NoStop}%
\bibitem [{\citenamefont {Abe}\ \emph {et~al.}(2001{\natexlab{a}})\citenamefont
  {Abe} \emph {et~al.}}]{SLD:2000ldu}%
  \BibitemOpen
  \bibfield  {author} {\bibinfo {author} {\bibfnamefont {K.}~\bibnamefont
  {Abe}} \emph {et~al.} (\bibinfo {collaboration} {SLD}),\ }\bibfield  {title}
  {\bibinfo {title} {{An improved direct measurement of leptonic coupling
  asymmetries with polarized Z bosons}},\ }\href@noop {} {\bibfield  {journal}
  {\bibinfo  {journal} {Phys. Rev. Lett.}\ }\textbf {\bibinfo {volume} {86}},\
  \bibinfo {pages} {1162} (\bibinfo {year} {2001}{\natexlab{a}})},\ \bibinfo
  {note} {leptonic event selection of the 2001 $A_{LR}$ analysis},\ \Eprint
  {https://arxiv.org/abs/hep-ex/0010015} {arXiv:hep-ex/0010015} \BibitemShut
  {NoStop}%
\bibitem [{\citenamefont {Meeus}(1998)}]{Meeus:1998}%
  \BibitemOpen
  \bibfield  {author} {\bibinfo {author} {\bibfnamefont {J.}~\bibnamefont
  {Meeus}},\ }\href@noop {} {\emph {\bibinfo {title} {Astronomical
  Algorithms}}},\ \bibinfo {edition} {2nd}\ ed.\ (\bibinfo  {publisher}
  {Willmann-Bell},\ \bibinfo {address} {Richmond},\ \bibinfo {year} {1998})\
  \bibinfo {note} {{GMST} polynomial used for the sidereal phase}\BibitemShut
  {NoStop}%
\bibitem [{\citenamefont {Friedsam}\ \emph {et~al.}(1985)\citenamefont
  {Friedsam}, \citenamefont {Goldsmith}, \citenamefont {Oren}, \citenamefont
  {Pietryka}, \citenamefont {Pitthan}, \citenamefont {Pushor},\ and\
  \citenamefont {Ruland}}]{Friedsam:1985cn316}%
  \BibitemOpen
  \bibfield  {author} {\bibinfo {author} {\bibfnamefont {H.}~\bibnamefont
  {Friedsam}}, \bibinfo {author} {\bibfnamefont {T.}~\bibnamefont {Goldsmith}},
  \bibinfo {author} {\bibfnamefont {W.}~\bibnamefont {Oren}}, \bibinfo {author}
  {\bibfnamefont {M.}~\bibnamefont {Pietryka}}, \bibinfo {author}
  {\bibfnamefont {R.}~\bibnamefont {Pitthan}}, \bibinfo {author} {\bibfnamefont
  {R.}~\bibnamefont {Pushor}},\ and\ \bibinfo {author} {\bibfnamefont
  {R.}~\bibnamefont {Ruland}},\ }\href@noop {} {\emph {\bibinfo {title} {{The
  SLC Vertical Survey Network}}}},\ \bibinfo {type} {Tech. Rep.}\ \bibinfo
  {number} {SLC Collider Note CN-316}\ (\bibinfo  {institution} {SLAC},\
  \bibinfo {year} {1985})\ \bibinfo {note} {in the documentation corpus of
  Ref.~\cite{zenodo:v2} (scan 0966); Fig.~1 is the as-built site map used for
  the beam azimuth}\BibitemShut {NoStop}%
\bibitem [{\citenamefont {Bong}\ \emph {et~al.}(2004)\citenamefont {Bong},
  \citenamefont {Emma}, \citenamefont {LeCocq}, \citenamefont {Montagne},\ and\
  \citenamefont {Welch}}]{LCLS:2003tn}%
  \BibitemOpen
  \bibfield  {author} {\bibinfo {author} {\bibfnamefont {E.}~\bibnamefont
  {Bong}}, \bibinfo {author} {\bibfnamefont {P.}~\bibnamefont {Emma}}, \bibinfo
  {author} {\bibfnamefont {C.}~\bibnamefont {LeCocq}}, \bibinfo {author}
  {\bibfnamefont {T.}~\bibnamefont {Montagne}},\ and\ \bibinfo {author}
  {\bibfnamefont {J.}~\bibnamefont {Welch}},\ }\href@noop {} {\emph {\bibinfo
  {title} {{LCLS Undulator Coordinate System}}}},\ \bibinfo {type} {Tech.
  Rep.}\ \bibinfo {number} {LCLS-TN-03-8, SLAC-TN-05-021}\ (\bibinfo
  {institution} {SLAC},\ \bibinfo {year} {2004})\ \bibinfo {note} {p.~2: linac
  azimuth $\alpha=261.45^{\circ}$, citing an unpublished Coast \& Geodetic
  Survey Geodesy Division note of Feb.~26, 1963}\BibitemShut {NoStop}%
\bibitem [{\citenamefont {Sheppard}(1985)}]{Sheppard:1985cn}%
  \BibitemOpen
  \bibfield  {author} {\bibinfo {author} {\bibfnamefont {J.~C.}\ \bibnamefont
  {Sheppard}},\ }\href@noop {} {\emph {\bibinfo {title} {{Required Energy Gain
  in the Linac}}}},\ \bibinfo {type} {Tech. Rep.}\ \bibinfo {number} {SLC
  Collider Memo CN-296}\ (\bibinfo  {institution} {SLAC},\ \bibinfo {year}
  {1985})\ \bibinfo {note} {table~1: as-built arc and final-focus bend budget;
  scan 0950 in the documentation corpus of Ref.~\cite{zenodo:v2}}\BibitemShut
  {NoStop}%
\bibitem [{\citenamefont {Apostolakis}\ \emph {et~al.}(1999)\citenamefont
  {Apostolakis} \emph {et~al.}}]{Apostolakis:1999cplear}%
  \BibitemOpen
  \bibfield  {author} {\bibinfo {author} {\bibfnamefont {A.}~\bibnamefont
  {Apostolakis}} \emph {et~al.} (\bibinfo {collaboration} {CPLEAR}),\
  }\bibfield  {title} {\bibinfo {title} {{Tests of the Equivalence Principle
  with Neutral Kaons}},\ }\href {https://doi.org/10.1016/S0370-2693(99)00271-3}
  {\bibfield  {journal} {\bibinfo  {journal} {Phys. Lett. B}\ }\textbf
  {\bibinfo {volume} {452}},\ \bibinfo {pages} {425} (\bibinfo {year}
  {1999})},\ \Eprint {https://arxiv.org/abs/hep-ex/9903005}
  {arXiv:hep-ex/9903005 [hep-ex]} \BibitemShut {NoStop}%
\bibitem [{\citenamefont {Kalaydzhyan}(2016)}]{Kalaydzhyan:2016positron}%
  \BibitemOpen
  \bibfield  {author} {\bibinfo {author} {\bibfnamefont {T.}~\bibnamefont
  {Kalaydzhyan}},\ }\bibfield  {title} {\bibinfo {title} {{Gravitational mass
  of positron from LEP synchrotron losses}},\ }\href
  {https://doi.org/10.1038/srep30461} {\bibfield  {journal} {\bibinfo
  {journal} {Sci. Rep.}\ }\textbf {\bibinfo {volume} {6}},\ \bibinfo {pages}
  {30461} (\bibinfo {year} {2016})},\ \Eprint
  {https://arxiv.org/abs/1508.04377} {arXiv:1508.04377 [hep-ph]} \BibitemShut
  {NoStop}%
\bibitem [{\citenamefont {Safronova}\ \emph {et~al.}(2018)\citenamefont
  {Safronova}, \citenamefont {Budker}, \citenamefont {DeMille}, \citenamefont
  {Jackson~Kimball}, \citenamefont {Derevianko},\ and\ \citenamefont
  {Clark}}]{Safronova:2018rev}%
  \BibitemOpen
  \bibfield  {author} {\bibinfo {author} {\bibfnamefont {M.~S.}\ \bibnamefont
  {Safronova}}, \bibinfo {author} {\bibfnamefont {D.}~\bibnamefont {Budker}},
  \bibinfo {author} {\bibfnamefont {D.}~\bibnamefont {DeMille}}, \bibinfo
  {author} {\bibfnamefont {D.~F.}\ \bibnamefont {Jackson~Kimball}}, \bibinfo
  {author} {\bibfnamefont {A.}~\bibnamefont {Derevianko}},\ and\ \bibinfo
  {author} {\bibfnamefont {C.~W.}\ \bibnamefont {Clark}},\ }\bibfield  {title}
  {\bibinfo {title} {{Search for New Physics with Atoms and Molecules}},\
  }\href {https://doi.org/10.1103/RevModPhys.90.025008} {\bibfield  {journal}
  {\bibinfo  {journal} {Rev. Mod. Phys.}\ }\textbf {\bibinfo {volume} {90}},\
  \bibinfo {pages} {025008} (\bibinfo {year} {2018})},\ \Eprint
  {https://arxiv.org/abs/1710.01833} {arXiv:1710.01833 [physics.atom-ph]}
  \BibitemShut {NoStop}%
\bibitem [{\citenamefont {Lange}\ \emph {et~al.}(2021)\citenamefont {Lange},
  \citenamefont {Huntemann}, \citenamefont {Rahm}, \citenamefont {Sanner},
  \citenamefont {Shao}, \citenamefont {Lipphardt}, \citenamefont {Tamm},
  \citenamefont {Weyers},\ and\ \citenamefont {Peik}}]{Lange:2021lpi}%
  \BibitemOpen
  \bibfield  {author} {\bibinfo {author} {\bibfnamefont {R.}~\bibnamefont
  {Lange}}, \bibinfo {author} {\bibfnamefont {N.}~\bibnamefont {Huntemann}},
  \bibinfo {author} {\bibfnamefont {J.~M.}\ \bibnamefont {Rahm}}, \bibinfo
  {author} {\bibfnamefont {C.}~\bibnamefont {Sanner}}, \bibinfo {author}
  {\bibfnamefont {H.}~\bibnamefont {Shao}}, \bibinfo {author} {\bibfnamefont
  {B.}~\bibnamefont {Lipphardt}}, \bibinfo {author} {\bibfnamefont
  {C.}~\bibnamefont {Tamm}}, \bibinfo {author} {\bibfnamefont {S.}~\bibnamefont
  {Weyers}},\ and\ \bibinfo {author} {\bibfnamefont {E.}~\bibnamefont {Peik}},\
  }\bibfield  {title} {\bibinfo {title} {{Improved limits for violations of
  local position invariance from atomic clock comparisons}},\ }\href
  {https://doi.org/10.1103/PhysRevLett.126.011102} {\bibfield  {journal}
  {\bibinfo  {journal} {Phys. Rev. Lett.}\ }\textbf {\bibinfo {volume} {126}},\
  \bibinfo {pages} {011102} (\bibinfo {year} {2021})},\ \Eprint
  {https://arxiv.org/abs/2010.06620} {arXiv:2010.06620 [physics.atom-ph]}
  \BibitemShut {NoStop}%
\bibitem [{\citenamefont {Kraiselburd}\ and\ \citenamefont
  {Vucetich}(2012)}]{Kraiselburd:2012eo}%
  \BibitemOpen
  \bibfield  {author} {\bibinfo {author} {\bibfnamefont {L.}~\bibnamefont
  {Kraiselburd}}\ and\ \bibinfo {author} {\bibfnamefont {H.}~\bibnamefont
  {Vucetich}},\ }\bibfield  {title} {\bibinfo {title} {{Constraining the
  fundamental interactions and couplings with E{\"o}tv{\"o}s experiments}},\
  }\href {https://doi.org/10.1016/j.physletb.2012.10.016} {\bibfield  {journal}
  {\bibinfo  {journal} {Phys. Lett. B}\ }\textbf {\bibinfo {volume} {718}},\
  \bibinfo {pages} {21} (\bibinfo {year} {2012})},\ \Eprint
  {https://arxiv.org/abs/1110.3527} {arXiv:1110.3527} \BibitemShut {NoStop}%
\bibitem [{\citenamefont {Kosteleck{\'y}}\ and\ \citenamefont
  {Tasson}(2011)}]{Kostelecky:2011matter}%
  \BibitemOpen
  \bibfield  {author} {\bibinfo {author} {\bibfnamefont {V.~A.}\ \bibnamefont
  {Kosteleck{\'y}}}\ and\ \bibinfo {author} {\bibfnamefont {J.~D.}\
  \bibnamefont {Tasson}},\ }\bibfield  {title} {\bibinfo {title}
  {{Matter-gravity couplings and Lorentz violation}},\ }\href
  {https://doi.org/10.1103/PhysRevD.83.016013} {\bibfield  {journal} {\bibinfo
  {journal} {Phys. Rev. D}\ }\textbf {\bibinfo {volume} {83}},\ \bibinfo
  {pages} {016013} (\bibinfo {year} {2011})},\ \Eprint
  {https://arxiv.org/abs/1006.4106} {arXiv:1006.4106 [gr-qc]} \BibitemShut
  {NoStop}%
\bibitem [{\citenamefont {Anderson}\ \emph {et~al.}(2004)\citenamefont
  {Anderson}, \citenamefont {Sher},\ and\ \citenamefont
  {Turan}}]{Anderson:2004qi}%
  \BibitemOpen
  \bibfield  {author} {\bibinfo {author} {\bibfnamefont {D.~L.}\ \bibnamefont
  {Anderson}}, \bibinfo {author} {\bibfnamefont {M.}~\bibnamefont {Sher}},\
  and\ \bibinfo {author} {\bibfnamefont {I.}~\bibnamefont {Turan}},\ }\bibfield
   {title} {\bibinfo {title} {{Lorentz and CPT violation in the Higgs
  sector}},\ }\href@noop {} {\bibfield  {journal} {\bibinfo  {journal} {Phys.
  Rev. D}\ }\textbf {\bibinfo {volume} {70}},\ \bibinfo {pages} {016001}
  (\bibinfo {year} {2004})},\ \Eprint {https://arxiv.org/abs/hep-ph/0403116}
  {arXiv:hep-ph/0403116} \BibitemShut {NoStop}%
\bibitem [{\citenamefont {Colladay}\ \emph {et~al.}(2017)\citenamefont
  {Colladay}, \citenamefont {Noordmans},\ and\ \citenamefont
  {Potting}}]{Colladay:2017zen}%
  \BibitemOpen
  \bibfield  {author} {\bibinfo {author} {\bibfnamefont {D.}~\bibnamefont
  {Colladay}}, \bibinfo {author} {\bibfnamefont {J.~P.}\ \bibnamefont
  {Noordmans}},\ and\ \bibinfo {author} {\bibfnamefont {R.}~\bibnamefont
  {Potting}},\ }\bibfield  {title} {\bibinfo {title} {{CPT and Lorentz
  Violation in the Photon and Z-Boson Sector}},\ }\href
  {https://doi.org/10.3390/sym9110248} {\bibfield  {journal} {\bibinfo
  {journal} {Symmetry}\ }\textbf {\bibinfo {volume} {9}},\ \bibinfo {pages}
  {248} (\bibinfo {year} {2017})},\ \Eprint {https://arxiv.org/abs/1711.03921}
  {arXiv:1711.03921 [hep-ph]} \BibitemShut {NoStop}%
\bibitem [{\citenamefont {Kepuladze}(2026)}]{Kepuladze:2026lzg}%
  \BibitemOpen
  \bibfield  {author} {\bibinfo {author} {\bibfnamefont {Z.}~\bibnamefont
  {Kepuladze}},\ }\bibfield  {title} {\bibinfo {title} {{Overlapping resonance
  branches of a gauge-invariant Lorentz-violating massive vector}},\
  }\href@noop {} {\  (\bibinfo {year} {2026})},\ \bibinfo {note} {preprint
  (submitted 1 August 2026)},\ \Eprint {https://arxiv.org/abs/2608.02655}
  {arXiv:2608.02655 [hep-ph]} \BibitemShut {NoStop}%
\bibitem [{\citenamefont {Ebrahimi}\ \emph {et~al.}(2025)\citenamefont
  {Ebrahimi}, \citenamefont {Moosavi~Nejad},\ and\ \citenamefont
  {Rezaei}}]{Ebrahimi:2025hjx}%
  \BibitemOpen
  \bibfield  {author} {\bibinfo {author} {\bibfnamefont {T.}~\bibnamefont
  {Ebrahimi}}, \bibinfo {author} {\bibfnamefont {S.~M.}\ \bibnamefont
  {Moosavi~Nejad}},\ and\ \bibinfo {author} {\bibfnamefont {Z.}~\bibnamefont
  {Rezaei}},\ }\bibfield  {title} {\bibinfo {title} {{Study of Lorentz
  invariance violation in polarized top quark decay}},\ }\href
  {https://doi.org/10.1140/epjc/s10052-025-14043-x} {\bibfield  {journal}
  {\bibinfo  {journal} {Eur. Phys. J. C}\ }\textbf {\bibinfo {volume} {85}},\
  \bibinfo {pages} {309} (\bibinfo {year} {2025})}\BibitemShut {NoStop}%
\bibitem [{\citenamefont {Rezaei}\ and\ \citenamefont
  {Moosavi~Nejad}(2025)}]{Rezaei:2025wiw}%
  \BibitemOpen
  \bibfield  {author} {\bibinfo {author} {\bibfnamefont {Z.}~\bibnamefont
  {Rezaei}}\ and\ \bibinfo {author} {\bibfnamefont {S.~M.}\ \bibnamefont
  {Moosavi~Nejad}},\ }\bibfield  {title} {\bibinfo {title} {{The effect of
  Lorentz violation on the azimuthal correlation contribution in polarized top
  quark decay}},\ }\href {https://doi.org/10.47176/ijpr.25.1.12046} {\bibfield
  {journal} {\bibinfo  {journal} {Iran. J. Phys. Res.}\ }\textbf {\bibinfo
  {volume} {25}},\ \bibinfo {pages} {119} (\bibinfo {year} {2025})},\ \bibinfo
  {note} {in Persian; English abstract at
  \url{https://ijpr.iut.ac.ir/article_3656.html}}\BibitemShut {NoStop}%
\bibitem [{\citenamefont {Abe}\ \emph {et~al.}(2001{\natexlab{b}})\citenamefont
  {Abe} \emph {et~al.}}]{SLD:2000ern}%
  \BibitemOpen
  \bibfield  {author} {\bibinfo {author} {\bibfnamefont {K.}~\bibnamefont
  {Abe}} \emph {et~al.} (\bibinfo {collaboration} {SLD}),\ }\bibfield  {title}
  {\bibinfo {title} {{First Symmetry Tests in Polarized $Z^0$ Decays to
  $b\bar{b}g$}},\ }\href {https://doi.org/10.1103/PhysRevLett.86.962}
  {\bibfield  {journal} {\bibinfo  {journal} {Phys. Rev. Lett.}\ }\textbf
  {\bibinfo {volume} {86}},\ \bibinfo {pages} {962} (\bibinfo {year}
  {2001}{\natexlab{b}})},\ \Eprint {https://arxiv.org/abs/hep-ex/0007051}
  {arXiv:hep-ex/0007051 [hep-ex]} \BibitemShut {NoStop}%
\bibitem [{\citenamefont {Burrows}\ and\ \citenamefont
  {Osland}(1997)}]{Burrows:1997dh}%
  \BibitemOpen
  \bibfield  {author} {\bibinfo {author} {\bibfnamefont {P.~N.}\ \bibnamefont
  {Burrows}}\ and\ \bibinfo {author} {\bibfnamefont {P.}~\bibnamefont
  {Osland}},\ }\bibfield  {title} {\bibinfo {title} {{Three-jet event
  orientation in $e^+e^-$ annihilation: new tests of the Standard Model}},\
  }\href {https://doi.org/10.1016/S0370-2693(97)00358-4} {\bibfield  {journal}
  {\bibinfo  {journal} {Phys. Lett. B}\ }\textbf {\bibinfo {volume} {400}},\
  \bibinfo {pages} {385} (\bibinfo {year} {1997})},\ \Eprint
  {https://arxiv.org/abs/hep-ph/9701424} {arXiv:hep-ph/9701424 [hep-ph]}
  \BibitemShut {NoStop}%
\bibitem [{\citenamefont {Accardi}\ \emph {et~al.}(2016)\citenamefont {Accardi}
  \emph {et~al.}}]{Accardi:2012qut}%
  \BibitemOpen
  \bibfield  {author} {\bibinfo {author} {\bibfnamefont {A.}~\bibnamefont
  {Accardi}} \emph {et~al.},\ }\bibfield  {title} {\bibinfo {title} {{Electron
  Ion Collider: The Next QCD Frontier}},\ }\href@noop {} {\bibfield  {journal}
  {\bibinfo  {journal} {Eur. Phys. J. A}\ }\textbf {\bibinfo {volume} {52}},\
  \bibinfo {pages} {268} (\bibinfo {year} {2016})},\ \Eprint
  {https://arxiv.org/abs/1212.1701} {arXiv:1212.1701} \BibitemShut {NoStop}%
\bibitem [{\citenamefont {Jejelava}\ and\ \citenamefont
  {Kepuladze}(2026)}]{Jejelava:2026liv}%
  \BibitemOpen
  \bibfield  {author} {\bibinfo {author} {\bibfnamefont {J.}~\bibnamefont
  {Jejelava}}\ and\ \bibinfo {author} {\bibfnamefont {Z.}~\bibnamefont
  {Kepuladze}},\ }\bibfield  {title} {\bibinfo {title} {{Probing Lorentz
  invariance violation in $Z$ boson mass measurements at high-energy
  colliders}},\ }\href@noop {} {\bibfield  {journal} {\bibinfo  {journal}
  {Phys. Rev. D}\ }\textbf {\bibinfo {volume} {113}},\ \bibinfo {pages}
  {075011} (\bibinfo {year} {2026})},\ \Eprint
  {https://arxiv.org/abs/2504.11248} {arXiv:2504.11248} \BibitemShut {NoStop}%
\end{thebibliography}%

\end{document}